\documentclass[a4paper]{article}

\usepackage{a4wide}
\usepackage[UKenglish]{babel}
\usepackage{graphicx,subfigure}
	\graphicspath{{./}{figures/}{figures/n1/}{figures/n2/}}
\usepackage[colorlinks=true,linkcolor=blue,citecolor=red]{hyperref}
\usepackage{xcolor}
\usepackage{amsmath,amsthm,bbold}
\usepackage{algorithm,algorithmic}
\usepackage[inline]{enumitem}
\usepackage{epstopdf}

\theoremstyle{definition}\newtheorem{assumption}{Assumption}[section]
\theoremstyle{remark}\newtheorem{remark}[assumption]{Remark}

\newcommand{\abs}[1]{\left\lvert#1\right\rvert}
\newcommand{\ave}[1]{\langle#1\rangle}

\newcommand{\N}{\mathbb{N}}
\renewcommand{\P}{\mathbb{P}}

\newcommand{\R}{\mathbb{R}}
\newcommand{\supp}{\operatorname{supp}}
\newcommand{\Var}{\operatorname{Var}}

\allowdisplaybreaks

\begin{document}
\title{Boltzmann-type description with cutoff \\ of Follow-the-Leader traffic models}

\author{	Andrea Tosin\thanks{Department of Mathematical Sciences ``G. L. Lagrange'', Politecnico di Torino, Italy} \and
		Mattia Zanella\thanks{Department of Mathematics ``F. Casorati'', University of Pavia, Italy}}
\date{}

\maketitle

\begin{abstract}
In this paper we consider a Boltzmann-type kinetic description of Follow-the-Leader traffic dynamics and we study the resulting asymptotic distributions, namely the counterpart of the Maxwellian distribution of the classical kinetic theory. In the Boltzmann-type equation we include a non-constant collision kernel, in the form of a cutoff, in order to exclude from the statistical model possibly unphysical interactions. In spite of the increased analytical difficulty caused by this further non-linearity, we show that a careful application of the quasi-invariant limit (an asymptotic procedure reminiscent of the grazing collision limit) successfully leads to a Fokker-Planck approximation of the original Boltzmann-type equation, whence stationary distributions can be explicitly computed. Our analytical results justify, from a genuinely model-based point of view, some empirical results found in the literature by interpolation of experimental data.

\medskip


\medskip

\end{abstract}

\section{Introduction}
Follow-the-Leader (FTL) traffic models are a class of microscopic models of vehicular traffic introduced in the fifties to describe the flow of vehicles along a one-directional road with no passing. Their basic assumption is that each vehicle adjusts its speed depending only on the speed of the vehicle ahead.

If the road is identified with the real axis and the position of the $i$th vehicle at time $t\geq 0$ is denoted by $x_i=x_i(t)\in\R$, a general FTL model is expressed by the following system of ordinary differential equations, cf.~\cite{gazis1961OR}:
\begin{equation}
	\begin{cases}
		\dot{x}_i=v_i \\
		\dot{v}_i=\dfrac{av_i^m}{{\left(x_{i+1}-x_i\right)}^n}\left(v_{i+1}-v_i\right),
	\end{cases}
	\qquad i=1,\,2,\,\dots,
	\label{eq:FTL}
\end{equation}
where $v_i=v_i(t)\in\R_+$ stands for the speed of the $i$th vehicle whereas $a\in\R_+$ and $m,\,n\in\N$ are parameters characterising the interaction of the $i$th vehicle with the ($i+1$)th vehicle ahead. In essence,~\eqref{eq:FTL} prescribes that the acceleration $\dot{v}_i$ is proportional to the relative speed of the two interacting vehicles through the non-constant factor
$$ \frac{av_i^m}{{\left(x_{i+1}-x_i\right)}^n}, $$
called the sensitivity of the driver.

In this paper, we will derive from~\eqref{eq:FTL} binary interaction rules on which we will ground a ``collisional'', viz. Boltzmann-type, kinetic description of traffic. Our ultimate goal is to deduce from the kinetic model the asymptotic distributions, i.e. the analogous of the Maxwellian distribution in classical gas dynamics, which depict several traffic features emerging at equilibrium. The latter include, for instance, the headway (sometimes also called clearance) and the time headway (sometimes simply referred to as the headway) statistical distributions, which in the transportation engineering literature are often estimated empirically and then interpolated by means of some known classes of probability density functions~\cite{abuelenin2015IEEE,jin2009TRC,yin2009IEEE}. By exploiting the renowned potential of classical methods of kinetic theory to deal with multi-agent systems~\cite{pareschi2013BOOK}, we will show that those statistical distributions can actually be obtained from a genuinely \textit{model-based} approach inspired by~\eqref{eq:FTL}. In our opinion, this constitutes both a further interesting validation of the microscopic model~\eqref{eq:FTL} and a contribution to a deeper understanding and interpretation of the empirical data beyond their interpolation.

As far as the advancement of kinetic methods for vehicular traffic is concerned, the contribution of this paper is twofold.

On one hand, we introduce kinetic traffic models based on binary interaction rules which are non-standard with respect to the mainstream in the reference literature and built on well consolidated microscopic traffic models. Virtually all kinetic models of traffic flow, from the pioneering ones~\cite{paveri1975TR,prigogine1971BOOK} to the most contemporary ones, see e.g.~\cite{delitala2007M3AS,herty2010KRM,klar1997JSP,puppo2017KRM,tosin2019MMS}, describe the microscopic state of the vehicles by means of their speed. Nevertheless, we show that if, rather than reinventing some \textit{ad hoc} though reasonable interaction rules, one wants to rely on the microscopic dynamics~\eqref{eq:FTL}, a more natural microscopic descriptor is the \textit{headway}
\begin{equation}
	s_i:=x_{i+1}-x_i,
	\label{eq:s}
\end{equation}
i.e. the space gap between a vehicle and the vehicle ahead. The advantage is that from the kinetic model one can then readily recover a statistical description of the traffic distributions mentioned before, which would instead be much less straightforward from a speed-based model.

On the other hand, we consider ``collisional'' models with \textit{cutoff}, which is a form of non-constant collision kernel quite rare in the kinetic literature of vehicular traffic and also, more in general, of multi-agent systems, see~\cite{cordier2005JSP,furioli2020M3AS,prigogine1971BOOK,tosin2019MCRF_preprint}. In particular, we prove that it is still possible to obtain a precise analytical characterisation of the asymptotic distributions in spite of the increased non-linearity of the Boltzmann-type equation caused by the non-constant kernel. It is worth anticipating that the introduction of a kinetic model with cutoff is not just a theoretical speculation. As it will be clear in the sequel, it is fundamental in order to ensure the physical consistency of the interaction schemes derived from~\eqref{eq:FTL}.

In more detail, the paper is organised as follows. In Section~\ref{sect:FTL_binary} we focus on the binary interaction schemes that may be derived from~\eqref{eq:FTL} for $m=n$ and we consider, in particular, those obtained for $n=1,\,2$, which will be relevant for the subsequent development of the theory. In Section~\ref{sect:Boltzmann} we introduce a Boltzmann-type kinetic model of the FTL dynamics based on the previous interaction rules and we show explicitly that a cutoff interaction kernel is needed, in general, to guarantee the physical consistency of the statistical description of the system. We anticipate that the role of such a kernel will be to exclude possible interactions leading to unphysical negative values of the post-interaction headway. In Section~\ref{sect:FP} we discuss the application of the asymptotic procedure called the \textit{quasi-invariant interaction limit} to our Boltzmann-type setting with cutoff. In particular we show that, in a suitable regime of the parameters of the binary interactions, it permits to recover a Fokker-Planck approximation of the original ``collisional'' equation, whence we compute explicitly the stationary distributions of the kinetic model. In Section~\ref{sect:numerics} we present some numerical tests which show that, consistently with the theoretical predictions in the appropriate regime of the microscopic parameters, the numerical solution of the Boltzmann-type equation approaches for large times the analytically computed stationary solution of the Fokker-Planck equation. Finally, in Section~\ref{sect:conclusions} we summarise the contents of the paper and we propose some concluding remarks.

\section{FTL-inspired binary interactions}
\label{sect:FTL_binary}
We observe that, using the headway~\eqref{eq:s}, we may rewrite model~\eqref{eq:FTL} in the form
$$ \frac{\dot{v}_i}{v_i^m}=a\frac{\dot{s}_i}{s_i^n}, \qquad i=1,\,2,\,\dots, $$
which allows for a direct integration of the $i$th equation depending on the values of the exponents $m$, $n$. Throughout the paper, we will focus in particular on the case $m=n$, which for $n=1$ gives
\begin{equation}
	v_i=Cs_i^a \qquad (C>0),
	\label{eq:vi-si.n=1}
\end{equation}
while for $n>1$ gives
\begin{equation}
	v_i=\frac{s_i}{{\left(a+Cs_i^{n-1}\right)}^\frac{1}{n-1}} \qquad (C\in\R).
	\label{eq:vi-si.n>1}
\end{equation}
In both cases, $C$ is an arbitrary integration constant. Since $s_i\in [0,\,+\infty)$ and $a>0$, we observe that in~\eqref{eq:vi-si.n=1} $v_i$ grows unboundedly for every $C>0$. Conversely, in~\eqref{eq:vi-si.n>1} $v_i$ increases from $0$ to $1/C^\frac{1}{n-1}$, which suggests to fix in this case $C=1$ so as to obtain a unitary maximum dimensionless speed of the vehicles.

\subsection{The case~\texorpdfstring{$\boldsymbol{n=1}$}{}}
\label{sect:n=1}
Writing~\eqref{eq:FTL} with $m=n=1$ for the $i$th and the ($i+1$)th vehicle, subtracting the corresponding equations and using~\eqref{eq:vi-si.n=1}, we determine the following equation for the headway $s_i$:
$$ \frac{d}{dt}\left[\dot{s}_i-C\left(s_{i+1}^a-s_i^a\right)\right]=0, $$
which implies
\begin{equation}
	\dot{s}_i=C\left(s_{i+1}^a-s_i^a\right)+c
	\label{eq:si.n=1}
\end{equation}
for an arbitrary integration constant $c\in\R$. We may fix $c$ by imposing, for instance, that the \textit{jammed traffic} state, namely the one with $s_i(t)=0$ for all $i=1,\,2,\,\dots$ and all $t\geq 0$, be a particular solution to this equation. Then $c=0$.

Having obtained a first order model, we are now in a position to apply the idea illustrated in~\cite{carrillo2010SIMA} to get a binary interaction rule: we approximate~\eqref{eq:si.n=1} in a short time interval of length $\Delta{t}>0$ (understood e.g., as the reaction time of the drivers) with the forward Euler formula, denoting $s:=s_i(t)$, $s_\ast:=s_{i+1}(t)$ and $s':=s_i(t+\Delta{t})$:
$$ s'=s+C\Delta{t}\left(s_\ast^a-s^a\right). $$
Since the ($i+1$)th vehicle does not modify instead its headway when interacting with the $i$th vehicle behind, the analogous binary rule for it reads simply $s_\ast'=s_\ast$.

In order to deal more realistically with partly random binary interactions, which model the non-deterministic aspects of driver behaviour, we further add to $s'$ a zero-mean stochastic fluctuation, which does not modify on average the main FTL dynamics. To this purpose, we introduce a random variable $\eta\in\R$ such that
\begin{equation}
	\ave{\eta}=0, \qquad \Var(\eta)=\ave{\eta^2}>0,
	\label{eq:eta}
\end{equation}
where $\ave{\cdot}$ denotes the expectation with respect to the law of $\eta$, and we finally write
\begin{equation}
	s'=s+\gamma\left(s_\ast^a-s^a\right)+s^\delta\eta, \qquad s_\ast'=s_\ast
	\label{eq:binary.n=1}
\end{equation}
with $\gamma:=C\Delta{t}>0$ for brevity. The coefficient $s^\delta$ with $\delta>0$ gives the intensity of the stochastic fluctuation. We assume that it increases with $s$, so that when a vehicle is close to the leading vehicle it mostly follows the deterministic FTL model. Conversely, when it is far from the leading vehicle it is mostly prone to the randomness of the driver behaviour.

\subsection{The case~\texorpdfstring{$\boldsymbol{n=2}$}{}}
For $n=2$, which here we regard as the prototype of the cases $n>1$, from~\eqref{eq:vi-si.n>1} we have
\begin{equation}
	v_i=\frac{s_i}{a+s_i}.
	\label{eq:vi-si.n=2}
\end{equation}
Proceeding like in Section~\ref{sect:n=1}, we determine now the following equation for the headway $s_i$:
$$ \frac{d}{dt}\left[\dot{s}_i-a\left(\frac{1}{a+s_i}-\frac{1}{a+s_{i+1}}\right)\right]=0, $$
namely
\begin{equation}
	\dot{s}_i=a\left(\frac{1}{a+s_i}-\frac{1}{a+s_{i+1}}\right)+c
	\label{eq:si.n=2}
\end{equation}
for an arbitrary integration constant $c\in\R$. In particular, we fix again $c=0$ in order for the jammed traffic state to be a solution also in this case.

A forward-in-time discretisation of~\eqref{eq:si.n=2} produces
$$ s'=s+a\Delta{t}\left(\frac{1}{a+s}-\frac{1}{a+s_\ast}\right). $$
Without loss of generality, here we may conveniently choose $\Delta{t}=\frac{\gamma}{a}$ for $\gamma>0$, as we anticipate that in this case we will be mainly interested in the regime of large $a$ (cf. Section~\ref{sect:FP.n=2}). Finally, adding a stochastic contribution to the interaction dynamics, we obtain the form of the binary interaction rules that we will consider in the sequel:
\begin{equation}
	s'=s+\gamma\left(\frac{1}{a+s}-\frac{1}{a+s_\ast}\right)+s^\delta\eta, \qquad s_\ast'=s_\ast,
	\label{eq:binary.n=2}
\end{equation}
where $\eta\in\R$ satisfies~\eqref{eq:eta} and $\delta>0$.

\section{Boltzmann-type kinetic description with cutoff}
\label{sect:Boltzmann}
Both interaction rules~\eqref{eq:binary.n=1},~\eqref{eq:binary.n=2} can be recast in the form
\begin{equation}\label{eq:interaction_micro}
\begin{split}
 s'&=s+I(s,\,s_\ast)+s^\delta\eta, \\
 s_\ast'&=s_\ast, 
\end{split}
\end{equation}
where the interaction function $I$ has the property that $I(s,\,s_\ast)=-I(s_\ast,\,s)$. In order to be physically admissible, these rules have to be such that $s',\,s_\ast'\geq 0$ for all $s,\,s_\ast\geq 0$, which is clearly obvious for $s_\ast'$ but not for $s'$.

In general, the possibility to guarantee $s'\geq 0$ depends strongly on $I$ and on the exponent $\delta$ of the coefficient of the stochastic fluctuation $\eta$. For instance, in the case~\eqref{eq:binary.n=2} with $\delta=1$ it can be proved that the conditions
$$ \eta\geq\frac{\gamma}{a^2}-1, \qquad \gamma<a^2 $$
are sufficient to ensure \textit{a priori} $s'\geq 0$ for all possible choices of $s,\,s_\ast\geq 0$, see~\cite{piccoli2019ZAMP_preprint} for the details. They amount to saying that the support of $\eta$ is bounded from the left, however in such a way that $\eta$ can take also negative values, which are essential in order to meet the requirements~\eqref{eq:eta}.

The same is instead not true if, for the same interaction rule~\eqref{eq:binary.n=2}, we consider e.g., $\delta=\frac{1}{2}$. Indeed, assume that we bound the support of $\eta$ from the left as $\eta\geq -\eta_0$ for some $0<\eta_0<+\infty$. Then, no matter how small $\eta_0$ is, if $\eta$ takes any negative value $\eta=\bar{\eta}\in [-\eta_0,\,0)$ and furthermore $s=\bar{\eta}^2$ we have
$$ s'=\gamma\left(\frac{1}{a+\bar{\eta}^2}-\frac{1}{a+s_\ast}\right), $$
thus every $s_\ast\in [0,\,\bar{\eta}^2)$ produces $s'<0$. A totally analogous situation occurs also for the interaction rule~\eqref{eq:binary.n=1} with $\delta=\frac{1}{2}$.

These examples demonstrate that, in general, not all the interactions modelled by~\eqref{eq:binary.n=1},~\eqref{eq:binary.n=2} are physically admissible. Those which are not have to be discarded from the statistical description of the system dynamics, in order to get the correct aggregate trends based only on the admissible interactions. This may be achieved by considering a Boltzmann-type description with cutoff:
\begin{equation}
	\frac{d}{dt}\int_{\R_+}\varphi(s)f(s,\,t)\,ds=\frac{1}{2\lambda}\int_{\R_+}\int_{\R_+}\ave{\chi(s'\geq 0)(\varphi(s')-\varphi(s))}f(s,\,t)f(s_\ast,\,t)\,ds\,ds_\ast,
	\label{eq:Boltzmann}
\end{equation}
where the kinetic distribution function $f=f(s,\,t):\R_+\times\R_+\to\R_+$ is such that $f(s,\,t)ds$ is the proportion of vehicles whose headway at time $t>0$ is comprised between $s$ and $s+ds$. Moreover, $\varphi:\R_+\to\R$ is an arbitrary observable quantity (test function) and, like before, $\ave{\cdot}$ denotes the expectation with respect to the law of $\eta$ contained in $s'$. The term
$$ \chi(s'\geq 0):=
	\begin{cases}
		1 & \text{if } s'\geq 0 \\
		0 & \text{otherwise}
	\end{cases} $$
plays the role of the cutoff (in particular, non-constant) collision kernel. Specifically, it discards the interactions producing $s'<0$, which in this way do not contribute to the evolution of $f$. Finally, the coefficient $\frac{1}{2\lambda}$ on the right-hand side comes from the general form of Boltzmann-type equations with non-symmetric interactions, cf.~\cite{pareschi2013BOOK}, the parameter $\lambda>0$ representing a relaxation time (in other words, $\frac{1}{\lambda}$ is the interaction frequency).

The presence of the non-constant collision kernel $\chi(s'\geq 0)$ makes it more difficult to extract from~\eqref{eq:Boltzmann} information on the aggregate trends of the system, such as e.g., the evolution of the statistical moments of the distribution function $f$:
$$ M_k(t):=\int_{\R_+}s^kf(s,\,t)\,ds \qquad (k\in\N). $$
Choosing $\varphi(s)=1$ in~\eqref{eq:Boltzmann} we obtain however
$$ \frac{d}{dt}\int_{\R_+}f(s,\,t)\,ds=0, $$
namely the conservation of the mass of the vehicles. This condition also implies that it is possible to understand $f$ as a probability density, up to possibly normalising it with respect to the constant total mass. 

Choosing instead $\varphi(s)=s$ in~\eqref{eq:Boltzmann} we discover
$$ \frac{dM_1}{dt}=\frac{1}{2\lambda}\int_{\R_+}\int_{\R_+}\ave{\chi(s'\geq 0)(I(s,\,s_\ast)+s^\delta\eta)}f(s,\,t)f(s_\ast,\,t)\,ds\,ds_\ast. $$
We notice that if the binary interactions are such that the condition $s'\geq 0$ may be guaranteed \textit{a priori}, like in the case~\eqref{eq:binary.n=2} with $\delta=1$, then $\chi(s'\geq 0)\equiv 1$ and
$$ \frac{dM_1}{dt}=\frac{1}{2\lambda}\int_{\R_+}\int_{\R_+}I(s,\,s_\ast)f(s,\,t)f(s_\ast,\,t)\,ds\,ds_\ast=0, $$
because $I$ is antisymmetric with respect to the line $s_\ast=s$. In this case, also the first moment of $f$, namely the mean headway of the vehicles, is conserved. However, this is in general not the case of the models that we are considering.

The difficulty to deal with the strongly non-linear Boltzmann-type equation~\eqref{eq:Boltzmann} may be bypassed in suitable asymptotic regimes, which allow one to transform~\eqref{eq:Boltzmann} in a kinetic model more amenable to analytical investigations. This does not only include the determination of the statistical moments $M_k$ but also the explicit computation of the stationary distribution, say $f^\infty=f^\infty(s)$, which in this context plays the role of the Maxwellian distribution of the classical kinetic theory in that it depicts the emerging trend when interactions are close to equilibrium.

\section{Fokker-Planck asymptotics}
\label{sect:FP}
An asymptotic regime in which a detailed study of a collisional kinetic model is often possible is that of the \textit{quasi-invariant interactions}, which has been introduced in~\cite{cordier2005JSP,toscani2006CMS} and is inspired by the grazing collision regime of the classical kinetic theory~\cite{villani1998PhD,villani1998ARMA}. The idea is to consider a regime of the parameters of the model in which each interaction produces a small variation of the microscopic state of the particles, so that a suitable approximation of the collision operator (right-hand side of~\eqref{eq:Boltzmann}) is possible. At the same time, in order to balance the little effect of the interactions and observe aggregate trends, it is necessary to increase correspondingly the interaction frequency, viz. to make the relaxation time $\lambda$ small.

We now illustrate in detail this procedure, which is very much inspired by~\cite{cordier2005JSP}, with reference to the interaction models introduced in Section~\ref{sect:FTL_binary}.

\subsection{The case~\texorpdfstring{$\boldsymbol{n=1}$}{}}
\label{sect:FP.n=1}
Let us consider model~\eqref{eq:binary.n=1} with $\delta=\frac{1}{2}$ and let us set\footnote{We choose $\lambda=\frac{\epsilon}{2}$ rather than $\lambda=\epsilon$ so as to absorb in the scaling the coefficient $\frac{1}{2}$ appearing in front of the collision operator in~\eqref{eq:Boltzmann}.}
\begin{equation}
	a=\Var(\eta)=\epsilon, \quad \lambda=\frac{\epsilon}{2}
	\label{eq:quasi-inv.n=1}
\end{equation}
where $0<\epsilon\ll 1$ is a parameter. Then the interactions are quasi-invariant, i.e. $s'\approx s$, because $s^\epsilon,\,s_\ast^\epsilon\approx 1$ and the distribution of $\eta$ is nearly the Dirac delta centred in zero. In particular, we can represent $\eta=\sqrt{\epsilon}Y$, where $Y$ is a random variable with zero mean and unitary variance. On the whole, the scaled interactions that we consider are
\begin{align*}
	s' &= s+\gamma\left(s_\ast^\epsilon-s^\epsilon\right)+\sqrt{\epsilon s}Y, \\
	s_\ast'&=s_\ast.
\end{align*}

The idea is now to manipulate the Boltzmann-type equation~\eqref{eq:Boltzmann} by taking advantage of the assumed smallness of $\epsilon$ and finally to approximate it, in the limit $\epsilon\to 0^+$, with a \textit{Fokker-Planck equation}. In the following, we will obtain such a limit equation in a formal fashion. Next, we will justify numerically our derivation by comparing the stationary solution of the obtained Fokker-Planck equation with the numerical solution to~\eqref{eq:Boltzmann} with $\epsilon$ small and $t$ large. For technical reasons, we will assume that:
\begin{assumption} \label{ass:logs.Y}
\begin{enumerate}[label=(\roman*)]
\item \label{ass.item:s.logs} $s,\,\log{s}\in L^p(\R_+;\,f(\cdot,\,t)ds)$ for some $p>0$ and all $t\geq 0$, i.e.:
$$ \int_{\R_+}s^pf(s,\,t)\,ds<+\infty, \quad \int_{\R_+}\abs{\log{s}}^pf(s,\,t)\,ds<+\infty \qquad \forall\,t\geq 0; $$
\item \label{ass.item:Y.symm} $Y$ is symmetric about $0$, i.e. $Y$ and $-Y$ have the same law;
\item \label{ass.item:Y.moments} $Y$ has bounded moments up to the order $3+\nu$ with $\nu>0$, i.e.
$$ \langle\abs{Y}^\alpha\rangle<+\infty \quad \text{for\ } 0\leq\alpha\leq 3+\nu. $$
\end{enumerate}
\end{assumption}

\begin{remark} \label{rem:logs.Y}
\begin{enumerate}[label=(\roman*)]
\item \label{rem.item:logs} Assumption~\ref{ass:logs.Y}\ref{ass.item:s.logs} implies, in particular, that $f$ has a minimum number of moments bounded. Moreover, it implies that $\log{s}\in L^{p'}(\R_+;\,f(\cdot,\,t)\,ds)$ for every $p'\in [0,\,p]$. Indeed, since $\abs{\log{s}}\geq 1$ for $s\in (0,\,e^{-1})\cup (e,\,+\infty)$, we have:
\begin{align*}
	\int_{\R_+}\abs{\log{s}}^{p'}f(s,\,t)\,ds &\leq \int_0^{\frac{1}{e}}\abs{\log{s}}^pf(s,\,t)\,ds+\int_{\frac{1}{e}}^{e}f(s,\,t)\,ds+\int_e^{+\infty}\abs{\log{s}}^pf(s,\,t)\,ds \\
	&\leq 1+\int_{\R_+}\abs{\log{s}}^pf(s,\,t)\,ds<+\infty.
\end{align*}
\item \label{rem.item:Y.symm} For every $a\geq 0$, Assumption~\ref{ass:logs.Y}\ref{ass.item:Y.symm} implies that $\P(Y<-a)=\P(Y>a)$, hence in particular that $\P(Y<-a)=\frac{1}{2}\P(\abs{Y}>a)$.
\end{enumerate}
\end{remark}

To begin with, we observe that $\chi(s'\geq 0)=1-\chi(s'<0)$, therefore we may rewrite~\eqref{eq:Boltzmann} as
\begin{align}
	\begin{aligned}[b]
		\frac{d}{dt}\int_{\R_+}\varphi(s)f(s,\,t)\,ds &= \frac{1}{\epsilon}\int_{\R_+}\int_{\R_+}\ave{\varphi(s')-\varphi(s)}f(s,\,t)f(s_\ast,\,t)\,ds\,ds_\ast \\
		&\phantom{=} -\frac{1}{\epsilon}\int_{\R_+}\int_{\R_+}\ave{\chi(s'<0)(\varphi(s')-\varphi(s))}f(s,\,t)f(s_\ast,\,t)\,ds\,ds_\ast \\
		&=: A_\epsilon(f,\,f)[\varphi](t)+R_\epsilon(f,\,f)[\varphi](t).
	\end{aligned}
	\label{eq:Boltzmann.Aeps_Reps}
\end{align}
Let now $\varphi\in C^\infty_c(\R_+)$. Since
\begin{equation}
	s_\ast^\epsilon-s^\epsilon=\epsilon\log{\frac{s_\ast}{s}}+\frac{1}{2}\epsilon^2\left(s_\ast^{\bar{\epsilon}}\log^2{s_\ast}-s^{\bar{\epsilon}}\log^2{s}\right) \qquad (\epsilon\to 0^+)
	\label{eq:s_eps.Taylor}
\end{equation}
with $\bar{\epsilon}\in (0,\,\epsilon)$ and since $s'<0$ is equivalent to
\begin{equation}
	Y<-\frac{s+\gamma(s_\ast^\epsilon-s^\epsilon)}{\sqrt{\epsilon s}}=:b_\epsilon(s,\,s_\ast),
	\label{eq:b_eps.n=1}
\end{equation}
by expanding $\varphi(s')-\varphi(s)$ in Taylor series around $s$ we get:
\begin{align}
	\begin{aligned}[b]
		\abs{R_\epsilon(f,\,f)[\varphi](t)}\leq\int_{\R_+}\int_{\R_+} &\left\langle\chi(Y<b_\epsilon(s,\,s_\ast))\left[\abs{\varphi'(s)}\left(\gamma\abs{\log{\frac{s_\ast}{s}}}
			+\sqrt{\frac{s}{\epsilon}}\abs{Y}+o(1)\right)\right.\right. \\
		&+ \frac{1}{2}\abs{\varphi''(s)}\left(2\sqrt{\epsilon}\gamma\abs{\log{\frac{s_\ast}{s}}}\sqrt{s}\abs{Y}+sY^2+o(\sqrt{\epsilon})\right) \\
		&+ \left.\left.\frac{1}{6}\abs{\varphi'''(\bar{s})}\left(\sqrt{\epsilon}s^{3/2}\abs{Y}^3+o(\sqrt{\epsilon})\right)\right]\right\rangle f(s,\,t)f(s_\ast,\,t)\,ds\,ds_\ast,
	\end{aligned}
	\label{eq:Reps}
\end{align}
where $\bar{s}\in (\min\{s,\,s_\ast\},\,\max\{s,\,s_\ast\})$. Using~\eqref{eq:s_eps.Taylor}, we see that the remainders $o(1)$, $o(\sqrt{\epsilon})$ denote terms which are bounded in $s,\,s_\ast$ because:
\begin{enumerate*}[label=(\roman*)]
\item $s$ is bounded away from $0$ and $+\infty$ thanks to the compactness of the support of $\varphi$ and all of its derivatives;
\item Assumption~\ref{ass:logs.Y}\ref{ass.item:s.logs} and Remark~\ref{rem:logs.Y}\ref{rem.item:logs} ensure the $f$-integrability of the powers of $s_\ast$ and $\abs{\log{s_\ast}}$, hence also of their products owing to H\"{o}lder's inequality, on $\R_+$ for $p$ sufficiently large.
\end{enumerate*}

The goal is now to take $\epsilon\to 0^+$ in~\eqref{eq:Reps}. Passing formally to the limit under the integrals, we have to handle expressions of the form $\ave{\abs{Y}^k\chi(Y<b_\epsilon(s,\,s_\ast))}$ for $k=0,\,\dots,\,3$. From H\"{o}lder's inequality we get
$$ \ave{\abs{Y}^k\chi(Y<b_\epsilon(s,\,s_\ast))}\leq\ave{\abs{Y}^{kq}}^\frac{1}{q}\ave{\chi(Y<b_\epsilon(s,\,s_\ast))^r}^\frac{1}{r}
	=\ave{\abs{Y}^{kq}}^\frac{1}{q}\P(Y<b_\epsilon(s,\,s_\ast))^\frac{1}{r}, $$
where $q,\,r\geq 1$ are such that $\frac{1}{q}+\frac{1}{r}=1$. Choosing $q\leq\frac{3+\nu}{k}$, in view of Assumption~\ref{ass:logs.Y}\ref{ass.item:Y.moments} we obtain $\ave{\abs{Y}^{kq}}<+\infty$ for every $k=0,\,\dots,\,3$. On the other hand, from the definition~\eqref{eq:b_eps.n=1} of $b_\epsilon(s,\,s_\ast)$ together with the expansion~\eqref{eq:s_eps.Taylor} we see that, for all fixed $s\in\supp{\varphi}$ and $s_\ast>0$, we can choose $\epsilon>0$ so small that $b_\epsilon(s,\,s_\ast)<0$. Consequently, owing to Assumption~\ref{ass:logs.Y}\ref{ass.item:Y.symm}, cf. also Remark~\ref{rem:logs.Y}\ref{rem.item:Y.symm}, and to Chebyshev's inequality\footnote{We recall that Chebyshev's inequality states that $\P(\abs{X-\mu}\geq k\sigma)\leq\frac{1}{k^2}$, where $X$ is a real-valued random variable with finite expectation $\mu$ and finite non-zero variance $\sigma^2$ and $k>0$. Here we apply it for $X=Y$, with $\mu=0$ and $\sigma^2=1$, and $k=\abs{b_\epsilon(s,\,s_\ast)}$.}, we have
$$ \P(Y<b_\epsilon(s,\,s_\ast))^\frac{1}{r}=\frac{1}{2^{1/r}}\P(\abs{Y}>\abs{b_\epsilon(s,\,s_\ast)})^\frac{1}{r}
	\leq\frac{1}{2^{1/r}b_\epsilon(s,\,s_\ast)^{2/r}}=\frac{(\epsilon s)^{1/r}}{2^{1/r}\left(s+\gamma\left(s_\ast^\epsilon-s^\epsilon\right)\right)^{2/r}}. $$
This shows that all the terms under the integrals in~\eqref{eq:Reps} tend pointwise to zero when $\epsilon\to 0^+$, including the one with $\sqrt{\frac{s}{\epsilon}}$ because $\sqrt{\epsilon}$ at the denominator can be compensated by the factor $\epsilon^{1/r}$ in the estimate above provided $r<2$. Consequently, we obtain
$$ R_\epsilon(f,\,f)[\varphi]\xrightarrow{\epsilon\to 0^+} 0. $$

Concerning the term $A_\epsilon(f,\,f)[\varphi]$, analogous calculations yield
\begin{align*}
	A_\epsilon(f,\,f)[\varphi](t) &= \int_{\R_+}\int_{\R_+}\varphi'(s)\left(\gamma\log{\frac{s_\ast}{s}}+o(1)\right)f(s,\,t)f(s_\ast,\,t)\,ds\,ds_\ast \\
	&\phantom{=} +\frac{1}{2}\int_{\R_+}\int_{\R_+}\varphi''(s)\left(s+\gamma^2\epsilon\log^2{\frac{s_\ast}{s}}+o(\epsilon)\right)f(s,\,t)f(s_\ast,\,t)\,ds\,ds_\ast \\
	&\phantom{=} +\frac{1}{6}\int_{\R_+}\int_{\R_+}\varphi'''(\bar{s})\left(\sqrt{\epsilon}s^{3/2}\ave{Y^3}+o(\sqrt{\epsilon})\right)f(s,\,t)f(s_\ast,\,t)\,ds\,ds_\ast,
\end{align*}
where we have taken into account that $\ave{Y}=0$, $\ave{Y^2}=1$. Using the compactness of $\supp{\varphi}$ and Assumption~\ref{ass:logs.Y}, we get then
$$ A_\epsilon(f,\,f)[\varphi](t)\xrightarrow{\epsilon\to 0^+}\int_{\R_+}\int_{\R_+}\left(\gamma\varphi'(s)\log{\frac{s_\ast}{s}}
	+\frac{1}{2}\varphi''(s)s\right)f(s,\,t)f(s_\ast,\,t)\,ds\,ds_\ast. $$

On the whole, in the limit $\epsilon\to 0^+$ we obtain from~\eqref{eq:Boltzmann.Aeps_Reps}
\begin{align}
	\begin{aligned}[b]
		\frac{d}{dt}\int_{\R_+}\varphi(s)f(s,\,t)\,ds &= \gamma\int_{\R_+}\varphi'(s)\left(\int_{\R_+}\log{s_\ast}f(s_\ast,\,t)\,ds_\ast-\log{s}\right)f(s,\,t)\,ds \\
		&\phantom{=} +\frac{1}{2}\int_{\R_+}\varphi''(s)sf(s,\,t)\,ds.
	\end{aligned}
	\label{eq:FP-weak.n=1}
\end{align}
If we denote
\begin{equation}
	L(t):=\int_{\R_+}\log{s_\ast}f(s_\ast,\,t)\,ds_\ast,
	\label{eq:L(t)}
\end{equation}
which is well defined in view of Assumption~\ref{ass:logs.Y}\ref{ass.item:s.logs}, integrating back by parts in~\eqref{eq:FP-weak.n=1} and using the arbitrariness of $\varphi\in C^\infty_c(\R_+)$ we recognise that $f$ satisfies the following Fokker-Planck equation in strong form with non-constant coefficients:
\begin{equation}
	\partial_tf=\frac{1}{2}\partial_s^2(sf)-\gamma\partial_s\left[(L(t)-\log{s})f\right].
	\label{eq:FP-strong.n=1}
\end{equation}
In summary,~\eqref{eq:FP-weak.n=1} and~\eqref{eq:FP-strong.n=1} represent the weak and the strong form of the asymptotic model which approximates~\eqref{eq:Boltzmann} in the quasi-invariant regime~\eqref{eq:quasi-inv.n=1} of the interactions~\eqref{eq:binary.n=1}.

Notice that, because of the compactness of $\supp{\varphi}$, the Fokker-Planck equation~\eqref{eq:FP-strong.n=1} comes without conditions at $s=0$ and $s\to +\infty$. Boundary conditions may be set by imposing, for instance, the fulfilment of some conservation properties. In particular, as it will be clear in a moment, in this context it is useful to guarantee that model~\eqref{eq:FP-strong.n=1} conserves in time the first moment of $f$, i.e. the mean headway of the vehicles. To study the evolution of $M_1$, we multiply~\eqref{eq:FP-strong.n=1} by $s$ and we integrate on $\R_+$. Recalling the definition~\eqref{eq:L(t)}, we discover:
$$ \frac{dM_1}{dt}=\left(\frac{1}{2}s^2\partial_sf(s,\,t)-\gamma L(t)sf(s,\,t)+\gamma s\log{s}f(s,\,t)\right\vert_0^{+\infty}, $$
therefore $M_1$ is conserved if, for all $t>0$, the terms $sf(s,\,t)$, $s^2\partial_sf(s,\,t)$ and $s\log{s}f(s,\,t)$ vanish when $s\to 0^+$ and $s\to +\infty$. Sufficient conditions for this are that, for all $t>0$, $f(s,\,t)$ and $\partial_sf(s,\,t)$ are bounded in $s=0$ and are infinitesimal of order greater than $2$ for $s\to +\infty$.

\begin{figure}[!t]
\centering
\includegraphics[width=0.8\textwidth]{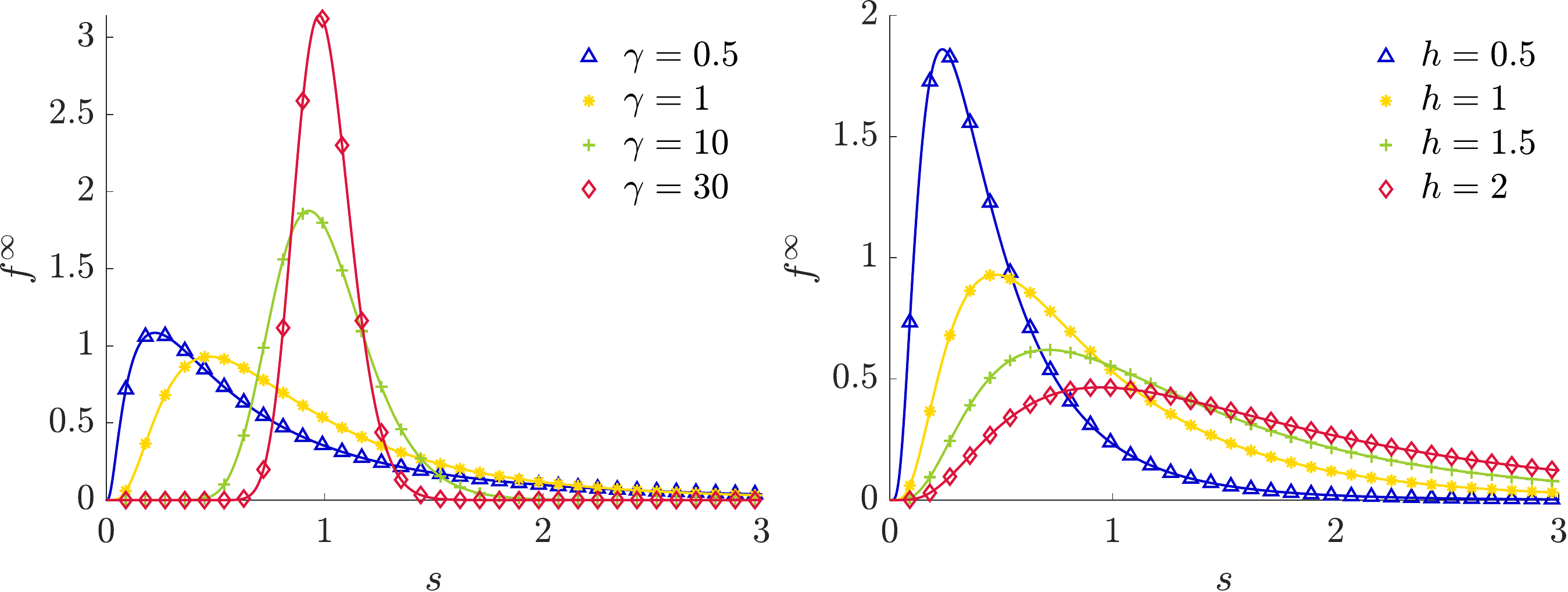}
\caption{The log-normal distribution~\eqref{eq:finf.n=1} predicted by model~\eqref{eq:binary.n=1} in the quasi-invariant regime~\eqref{eq:quasi-inv.n=1} for: $h=1$ and various $\gamma>0$ (\textbf{left}); $\gamma=1$ and various $h>0$ (\textbf{right}).}
\label{fig:finf_lognormal}
\end{figure}

Next, we may use~\eqref{eq:FP-strong.n=1} to obtain the stationary distribution $f^\infty$, which satisfies
$$ \frac{1}{2}\partial_s(sf^\infty)-\gamma(L^\infty-\log{s})f^\infty=0, $$
where $L^\infty:=\lim_{t\to+\infty}L(t)$ is so far unknown. This differential equation can be easily solved by separation of variables. Its unique solution with unitary mass is the function
$$ f^\infty(s)=\frac{\sqrt{\gamma}}{s\sqrt{\pi}}e^{-\gamma{\left(\log{s}-L^\infty\right)}^2}, $$
namely a \textit{log-normal} probability density function with parameters $L^\infty\in\R$ and $\frac{1}{\sqrt{2\gamma}}>0$. Notice that such an $f^\infty$ satisfies the boundary conditions stated above. From the known formulas of the moments of a log-normally distributed random variable we deduce, in particular, that the mean of $f^\infty$ is
$$ M_1^\infty:=\int_{\R_+}sf^\infty(s)\,ds=e^{L^\infty+\frac{1}{4\gamma}}, $$
which, owing to the conservation in time of $M_1$, has to coincide with the constant mean headway of the system, say $h>0$. Therefore we can express $L^\infty=\log{h}-\frac{1}{4\gamma}$ and finally write
\begin{equation}
	f^\infty(s)=\frac{\sqrt{\gamma}}{s\sqrt{\pi}}e^{-\gamma{\left[\log{s}-\left(\log{h}-\frac{1}{4\gamma}\right)\right]}^2},
	\label{eq:finf.n=1}
\end{equation}
see Figure~\ref{fig:finf_lognormal}.

In the transportation engineering literature, the log-normal distribution has often been reported to fit well the empirical data of vehicle interspacings, see e.g.,~\cite{jin2009TRC,panichpapiboon2013IEEE}. This motivated some attempts to justify, either analytically or computationally, the emergence of the log-normal distribution using particle models of traffic, which however rely often on case-specific assumptions~\cite{greenberg1966ARR,jin2009TRC}. Recently, a much more limpid theoretical explanation of the emergence of the log-normal distribution from microscopic agent dynamics has been provided in~\cite{gualandi2019M3AS} using kinetic theory methods which also inspire the present work. Nevertheless, in~\cite{gualandi2019M3AS} the authors do not consider actual interactions among the agents; rather, they assume that the agents change independently their state, trying to approach a recommended optimal state. On the basis of the prospect theory by Kahneman and Tversky~\cite{kahneman1979ECON}, such a change is assumed to require an asymmetric effort, depending on whether the current state is above or below the optimal one. It is then such an asymmetry which generates the log-normal distribution. In~\cite{gualandi2019M3AS} the authors recast vehicular traffic in this conceptual scheme by assuming that each driver adjusts the distance $s$ from the leading vehicle aiming at an optimal headway $\bar{s}$. The asymmetric effort depends on the fact that it should be easier to approach the optimal headway from above, i.e. for $s>\bar{s}$, because this corresponds to accelerating to get closer to the leading vehicle; while it should be harder to approach it from below, i.e. for $s<\bar{s}$, because this corresponds to braking to get farther from the leading vehicle. While certainly reasonable and embraceable, unlike~\eqref{eq:binary.n=1} such a behavioural model is not grounded on existing particle descriptions of traffic acknowledged in the literature. Our contribution has instead the merit to show that the log-normal distribution~\eqref{eq:finf.n=1} can be obtained organically from true binary interactions motivated by well consolidated microscopic traffic models.

Recalling~\eqref{eq:vi-si.n=1}, we also deduce the following relationship between the time headway $\tau$ and the headway $s$:
\begin{equation}
	\tau:=\frac{s}{v}=\frac{s^{1-a}}{C}.
	\label{eq:tau.n=1}
\end{equation}
Without loss of generality, let us fix $C=1$. If, consistently with the quasi-invariant regime~\eqref{eq:quasi-inv.n=1}, we assume that $a$ is small, in particular $a<1$, we can use the distribution~\eqref{eq:finf.n=1} together with the transformation~\eqref{eq:tau.n=1} to obtain the stationary distribution $g^\infty=g^\infty(\tau)$ of the time headway:
\begin{align*}
	g^\infty(\tau) &= \frac{1}{1-a}\tau^\frac{a}{1-a}f^\infty(\tau^{1/(1-a)}) \\
	&= \frac{\sqrt{\gamma}}{\tau(1-a)\sqrt{\pi}}e^{-\frac{\gamma}{(1-a)^2}\left[\log{\tau}-(1-a)\left(\log{h}-\frac{1}{4\gamma}\right)\right]^2},
\end{align*}
namely in turn a log-normal probability density function. The experimental literature widely acknowledges that the measured time headways distribute, with good approximation, according to a log-normal profile, see e.g.,~\cite{chen2010IEEE,wisitpongphan2007IEEE} and references therein. Also in this case, \textit{ad hoc} particle models have already been proposed~\cite{chen2010IEEE} to justify the emergence of such a distribution. Nevertheless, we believe that the kinetic approach presented here offers a more general and organic explanation grounded on simpler and sounder first principles.

Finally, from~\eqref{eq:finf.n=1} and the transformation~\eqref{eq:vi-si.n=1} with $C=1$ we derive the stationary distribution $k^\infty=k^\infty(v)$ of the speed $v$ in the quasi-invariant limit~\eqref{eq:quasi-inv.n=1}, i.e. in particular for $a$ small:
\begin{align*}
	k^\infty(v) &= \frac{1}{a}v^\frac{1-a}{a}f^\infty(v^{1/a}) \\
	&= \frac{\sqrt{\gamma}}{va\sqrt{\pi}}e^{-\frac{\gamma}{a^2}\left[\log{v}-a\left(\log{h}-\frac{1}{4\gamma}\right)\right]^2}.
\end{align*}
We observe that this is again a log-normal probability density function, hence it has in particular a slim tail for $v\to +\infty$. This partially mitigates the drawback of the unbounded speed allowed by the relationship~\eqref{eq:vi-si.n=1} because it implies that, at least in the quasi-invariant regime~\eqref{eq:quasi-inv.n=1}, very high speed values are quite rarely produced by the microscopic interaction model. In particular, the mean speed is $h^ae^{\frac{a}{2\gamma}\left(a-\frac{1}{2}\right)}$. Interestingly, in~\cite{ni2018AMM} the authors suggest that a log-normal profile may provide an acceptable fitting of the experimental speed distribution, at least as far as the empirical data used in their study are concerned.

\subsection{The case~\texorpdfstring{$\boldsymbol{n=2}$}{}}
\label{sect:FP.n=2}
We now consider model~\eqref{eq:binary.n=2} with $\delta=\frac{1}{2}$ and we focus on the following regime of the parameters:
\begin{equation}
	a=\frac{1}{\sqrt{\epsilon}}, \quad \Var(\eta)=\epsilon, \quad \lambda=\frac{\epsilon}{2},
	\label{eq:quasi-inv.n=2}
\end{equation}
with $0<\epsilon\ll 1$ as usual. The scaled interaction rules take then the form
\begin{align*}
	s' &= s+\gamma\epsilon\frac{s_\ast-s}{(1+\sqrt{\epsilon}s)(1+\sqrt{\epsilon}s_\ast)}+\sqrt{\epsilon s}Y, \\
	s_\ast' &= s_\ast,
\end{align*}
whence we see that they are quasi-invariant because $s'\approx s$ for $\epsilon$ small.

To obtain from~\eqref{eq:Boltzmann} the Fokker-Planck equation in the quasi-invariant limit we proceed along the lines of Section~\ref{sect:FP.n=1}, requiring in particular the validity of Assumption~\ref{ass:logs.Y} except for the integrability of $\log{s}$ claimed at point~\ref{ass.item:s.logs}.

After rewriting~\eqref{eq:Boltzmann} in the form~\eqref{eq:Boltzmann.Aeps_Reps}, we observe that $s'<0$ implies
\begin{equation}
	Y<-\frac{1}{\sqrt{\epsilon s}}\left(s+\gamma\epsilon\frac{s_\ast-s}{(1+\sqrt{\epsilon}s)(1+\sqrt{\epsilon}s_\ast)}\right)
		\leq\frac{\gamma\epsilon-1}{\sqrt{\epsilon}}\sqrt{s}=:b_\epsilon(s),
	\label{eq:b_eps.n=2}
\end{equation}
whence $\chi(s'<0)\leq\chi(Y<b_\epsilon(s))$. Moreover, $\abs{\frac{s_\ast-s}{(1+\sqrt{\epsilon}s)(1+\sqrt{\epsilon}s_\ast)}}\leq\abs{s_\ast-s}$. Thus, for $\varphi\in C^\infty_c(\R_+)$ we estimate:
\begin{align*}
	\abs{R_\epsilon(f,\,f)[\varphi](t)}\leq\int_{\R_+}\int_{\R_+} &\biggl\langle\chi(Y<b_\epsilon(s))\biggl[\abs{\varphi'(s)}\left(\gamma\abs{s_\ast-s}+\sqrt{\frac{s}{\epsilon}}\abs{Y}\right) \\
	&+ \frac{1}{2}\abs{\varphi''(s)}\left(\gamma^2\epsilon(s_\ast-s)^2+2\gamma\sqrt{\epsilon s}\abs{s_\ast-s}\abs{Y}+sY^2\right) \\
	&+ \frac{1}{6}\abs{\varphi'''(\bar{s})}
		\begin{aligned}[t]
			&\left(\gamma^3\epsilon^2\abs{s_\ast-s}^3+3\gamma\epsilon\sqrt{\epsilon s}(s_\ast-s)^2\abs{Y}\right. \\
			&\left. +3\gamma\epsilon s\abs{s_\ast-s}Y^2+\sqrt{\epsilon}s^{3/2}\abs{Y}^3\right)\biggl]\biggl\rangle f(s,\,t)f(s_\ast,\,t)\,ds\,ds_\ast.
		\end{aligned}
\end{align*}
To manipulate the terms $\ave{\abs{Y}^k\chi(Y<b_\epsilon(s))}$, $k=0,\,\dots,\,3$, we resort again to H\"{o}lder's inequality:
$$ \ave{\abs{Y}^k\chi(Y<b_\epsilon)}\leq\ave{\abs{Y}^{kq}}^\frac{1}{q}\ave{\chi(Y<b_\epsilon(s))^r}^\frac{1}{r}
	=\ave{\abs{Y}^{kq}}^\frac{1}{q}\P(Y<b_\epsilon(s))^\frac{1}{r}, $$
where $q,\,r\geq 1$ are chosen like in Section~\ref{sect:FP.n=1}. In view of Assumption~\ref{ass:logs.Y}\ref{ass.item:Y.moments}, it results $\ave{\abs{Y}^{kq}}<+\infty$ for $k=0,\,\dots,\,3$. Furthermore, from~\eqref{eq:b_eps.n=2} we see that we can take $\epsilon$ so small, in particular $\epsilon<\frac{1}{\gamma}$, that $b_\epsilon(s)<0$ for all $s>0$. Consequently, invoking Assumption~\ref{ass:logs.Y}\ref{ass.item:Y.symm} and Remark~\ref{rem:logs.Y}\ref{rem.item:Y.symm} together with Chebyshev's inequality, we obtain
$$ \P(Y<b_\epsilon(s))^\frac{1}{r}=\frac{1}{2^{1/r}}\P(\abs{Y}>\abs{b_\epsilon(s)})^\frac{1}{r}
	\leq\frac{1}{2^{1/r}b_\epsilon(s)^{2/r}}=\frac{\epsilon^{1/r}}{(2s)^{1/r}(\gamma\epsilon-1)^{2/r}}. $$
Plugging this into the estimate of $\abs{R_\epsilon(f,\,f)[\varphi](t)}$, and recalling that $s\in\supp{\varphi}$ is bounded away from $0,\,+\infty$ while the powers of $s_\ast$ are $f$-integrable thanks to Assumption~\ref{ass:logs.Y}\ref{ass.item:s.logs} with $p$ sufficiently large, we conclude
$$ R_\epsilon(f,\,f)[\varphi](t)\xrightarrow{\epsilon\to 0^+}0. $$
In particular, we stress that the term containing $\sqrt{\frac{s}{\epsilon}}$ vanishes in the limit because $\sqrt{\epsilon}$ at the denominator is compensated by the factor $\epsilon^{1/r}$ with $r<2$.

Concerning the term $A_\epsilon(f,\,f)[\varphi](t)$, by means of analogous calculations and taking into account that $\ave{Y}=0$, $\ave{Y^2}=1$ and that $\ave{\abs{Y}^3}<+\infty$, cf. Assumption~\ref{ass:logs.Y}\ref{ass.item:Y.moments}, we find:
\begin{align*}
	A_\epsilon(f,\,f)[\varphi](t) &= \gamma\int_{\R_+}\int_{\R_+}\varphi'(s)\frac{s_\ast-s}{(1+\sqrt{\epsilon}s)(1+\sqrt{\epsilon}s_\ast)}f(s,\,t)f(s_\ast,\,t)\,ds\,ds_\ast \\
	&\phantom{=} +\frac{1}{2}\int_{\R_+}\int_{\R_+}\varphi''(s)\left(\frac{\gamma^2\epsilon(s_\ast-s)^2}{(1+\sqrt{\epsilon}s)^2(1+\sqrt{\epsilon}s_\ast)^2}+s\right)f(s,\,t)f(s_\ast,\,t)\,ds\,ds_\ast \\
	&\phantom{=} +\frac{1}{6}\int_{\R_+}\int_{\R_+}\varphi'''(\bar{s})
		\begin{aligned}[t]
			&\biggl(\frac{\gamma^3\epsilon^2(s_\ast-s)^3}{(1+\sqrt{\epsilon}s)^3(1+\sqrt{\epsilon}s_\ast)^3}+\frac{3\gamma\epsilon(s_\ast-s)s}{(1+\sqrt{\epsilon}s)(1+\sqrt{\epsilon}s_\ast)} \\
			&+ \sqrt{\epsilon}s^{3/2}\ave{Y^3}\biggr)f(s,\,t)f(s_\ast,\,t)\,ds\,ds_\ast
		\end{aligned} \\
	&\xrightarrow{\epsilon\to 0^+}\int_{\R_+}\int_{\R_+}\left(\gamma\varphi'(s)(s_\ast-s)+\frac{1}{2}\varphi''(s)s\right)f(s,\,t)f(s_\ast,\,t)\,ds\,ds_\ast,
\end{align*}
hence for $\epsilon\to 0^+$ we finally get from~\eqref{eq:Boltzmann.Aeps_Reps}
$$ \frac{d}{dt}\int_{\R_+}\varphi(s)f(s,\,t)\,ds=\gamma\int_{\R_+}\varphi'(s)\left(M_1(t)-s\right)f(s,\,t)\,ds
	+\frac{1}{2}\int_{\R_+}\varphi''(s)sf(s,\,t)\,ds. $$

Integrating back by parts and invoking the arbitrariness of $\varphi\in C^\infty_c(\R_+)$, we deduce that $f$ satisfies the Fokker-Planck equation
\begin{equation}
	\partial_tf=\frac{1}{2}\partial_s^2(sf)-\gamma\partial_s((M_1(t)-s)f),
	\label{eq:FP-strong.n=2}
\end{equation}
which comes again without conditions at $s=0$ and for $s\to +\infty$ because of the compactness of $\supp{\varphi}$. Like in Section~\ref{sect:FP.n=1}, it is convenient to fix these conditions in such a way that $M_1$ is conserved in time. To this purpose, we multiply~\eqref{eq:FP-strong.n=2} by $s$ and we integrate on $\R_+$ to discover:
$$ \frac{dM_1}{dt}=\left(\frac{1}{2}s^2\partial_sf(s,\,t)-\gamma M_1(t)sf(s,\,t)+\gamma s^2f(s,\,t)\right\vert_0^{+\infty}. $$
From here we see that, analogously to Section~\ref{sect:FP.n=1}, sufficient conditions for $\frac{dM_1}{dt}=0$ are the fact that, for all $t>0$, $f(s,\,t)$ and $\partial_sf(s,\,t)$ are bounded at $s=0$ and be infinitesimal of order greater than $2$ for $s\to +\infty$.

\begin{figure}[!t]
\centering
\includegraphics[width=0.8\textwidth]{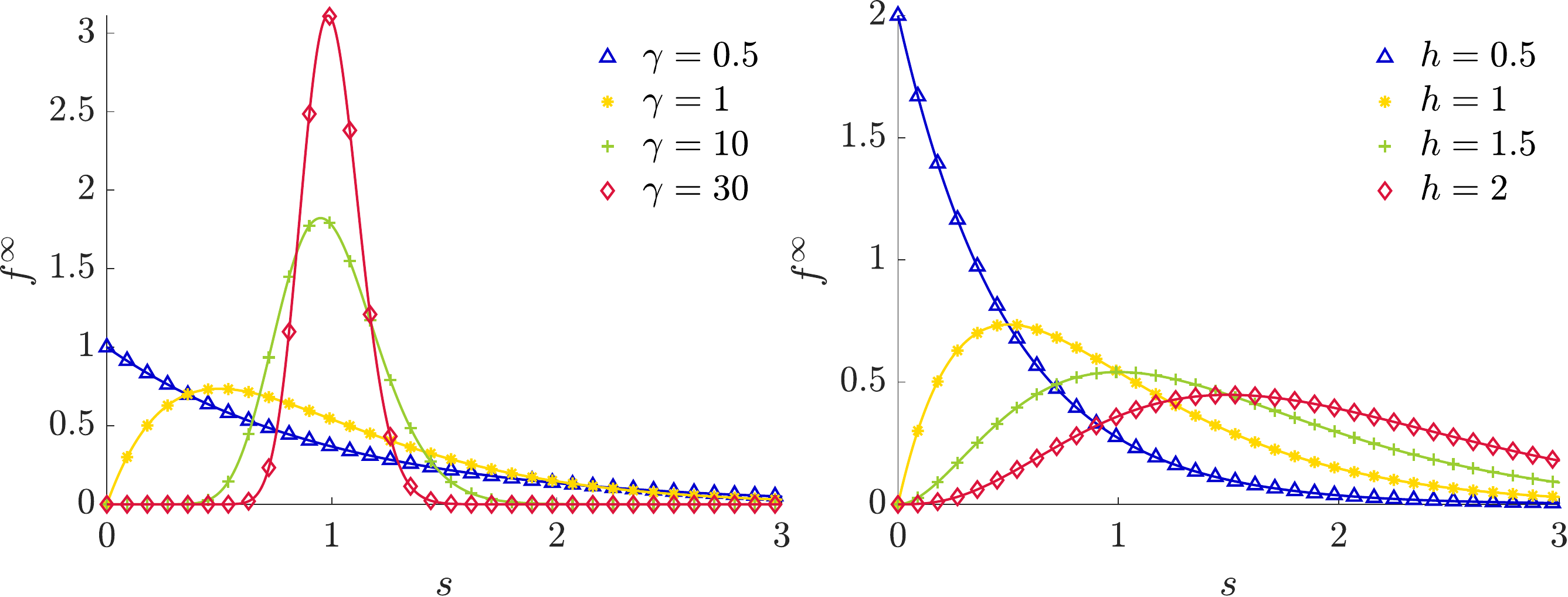}
\caption{The gamma distribution~\eqref{eq:finf.n=2} predicted by model~\eqref{eq:binary.n=2} in the quasi-invariant regime~\eqref{eq:quasi-inv.n=2} for: $h=1$ and various $\gamma>0$ (\textbf{left}); $\gamma=1$ and various $h>0$ (\textbf{right}).}
\label{fig:finf_gamma}
\end{figure}

Under such conditions we can set $M_1(t)=h$ for all $t\geq 0$, so that from~\eqref{eq:FP-strong.n=2} we obtain in particular the following unique stationary distribution with unitary mass:
\begin{equation}
	f^\infty(s)=\frac{(2\gamma)^{2\gamma h}}{\Gamma(2\gamma h)}s^{2\gamma h-1}e^{-2\gamma s},
	\label{eq:finf.n=2}	
\end{equation}
namely a \textit{gamma} probability density function with shape parameter $2\gamma h>0$ and rate parameter $2\gamma>0$, see Figure~\ref{fig:finf_gamma}.

In the transportation engineering literature, also the gamma distribution is sometimes used to fit the experimental measurements of the vehicle interspacings, see e.g.,~\cite{cowan1975TR}. Our derivation demonstrates that it may be justified out of Follow-the-Leader microscopic dynamics~\eqref{eq:FTL} with an appropriate choice of the exponents $m$, $n$.

\begin{figure}[!t]
\centering
\includegraphics[width=0.8\textwidth]{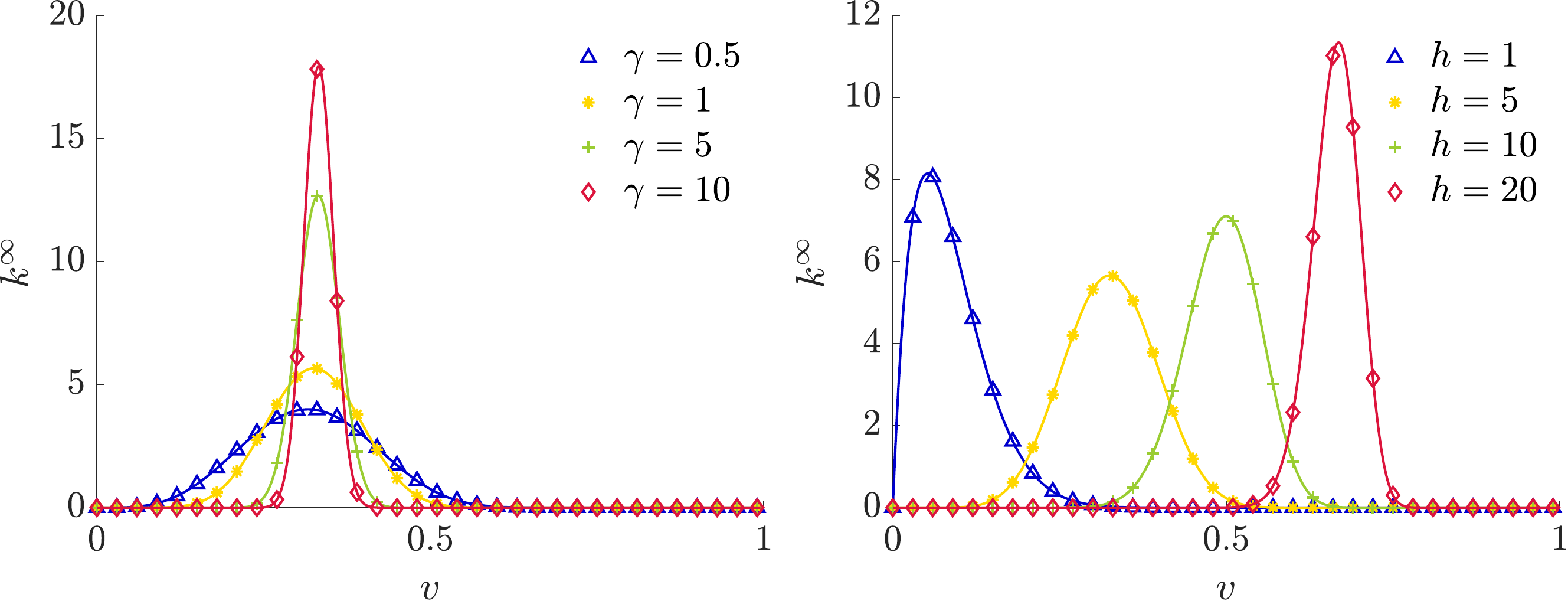}
\caption{The speed distribution~\eqref{eq:kinf.n=2} with $a=10$ for: $h=5$ and various $\gamma>0$ (\textbf{left}); $\gamma=1$ and various $h>0$ (\textbf{right}).}
\label{fig:kinf_gamma}
\end{figure}

Recalling~\eqref{eq:vi-si.n=2}, we see that the time headway is simply
$$ \tau=\frac{s}{v}=a+s, $$
hence its asymptotic distribution $g^\infty$, which is supported in the interval $[a,\,+\infty)$ because $s\geq 0$ implies now $\tau\geq a$, is obtained by translating $f^\infty$ rightward:
$$ g^\infty(\tau)=f^\infty(\tau-a)\chi(\tau\geq a). $$
Instead, the asymptotic distribution $k^\infty$ of $v$ resulting from the transformation~\eqref{eq:vi-si.n=2} reads
\begin{align}
	\begin{aligned}[b]
		k^\infty(v) &= \frac{a}{(1-v)^2}f^\infty\left(\frac{av}{1-v}\right) \\
		&= \frac{(2\gamma a)^{2\gamma h}}{\Gamma(2\gamma h)}\cdot\frac{v^{2\gamma h-1}}{(1-v)^{2\gamma h+1}}e^{-2\gamma a\frac{v}{1-v}}
	\end{aligned}
	\label{eq:kinf.n=2}
\end{align}
and is naturally supported in $[0,\,1]$, see Figure~\ref{fig:kinf_gamma}. Notice that for $v\to 1^-$ we have $k^\infty(v)\to 0$. Conversely, for $v\to 0^+$ we may have $k^\infty(v)\to 0$ if $2\gamma h>1$; $k^\infty(v)\to (2\gamma a)^{2\gamma h}/\Gamma(2\gamma h)$ if $2\gamma h=1$; or $k^\infty(v)\to +\infty$ if $2\gamma h<1$. In the latter case, the singularity of $k^\infty$ at $v=0$ is however integrable.

We stress that, consistently with the quasi-invariant regime~\eqref{eq:quasi-inv.n=2} motivating the form~\eqref{eq:finf.n=2} of $f^\infty$, in both expressions of $g^\infty$ and $k^\infty$ the parameter $a$ has to be understood as sufficiently large.

\subsubsection{The case~\texorpdfstring{$\boldsymbol{\delta=1}$}{}}
If we consider model~\eqref{eq:binary.n=2} with $\delta=1$ then, owing to the discussion at the beginning of Section~\ref{sect:Boltzmann}, we can guarantee \textit{a priori} the fulfilment of the condition $s'\geq 0$ for all $s,\,s_\ast\geq 0$ with an appropriate choice of the parameters $a$, $\gamma$ and of the random variable $\eta$. This implies that $\chi(s'\geq 0)\equiv 1$ in~\eqref{eq:Boltzmann}, hence, under the same scaling~\eqref{eq:quasi-inv.n=2}, the quasi-invariant limit simplifies considerably (it basically requires to deal only with the term $A_\epsilon(f,\,f)[\varphi]$) and yields finally the Fokker-Planck equation
$$ \partial_tf=\frac{1}{2}\partial^2_s(s^2f)-\gamma\partial_s((h-s)f), $$
which differs from~\eqref{eq:FP-strong.n=2} only in the coefficient of $f$ in the second order derivative. The unique stationary solution with unitary mass is now
$$ f^\infty(s)=\frac{(2\gamma h)^{1+2\gamma}}{\Gamma(1+2\gamma)}\cdot\frac{e^{-\frac{2\gamma h}{s}}}{s^{2(1+\gamma)}}, $$
namely an \textit{inverse gamma} probability density function with shape parameter $1+2\gamma>0$ and scale parameter $2\gamma h>0$. Unlike the stationary distributions~\eqref{eq:finf.n=1},~\eqref{eq:finf.n=2}, this $f^\infty$ features a \textit{fat tail}, indeed it behaves like $s^{-2(1+\gamma)}$ for $s\to +\infty$. Interestingly, fat tailed headway distributions are also reported in the experimental literature~\cite{abuelenin2015IEEE} and justified with the presence of high occupancy vehicles in the traffic stream.

\section{Numerical tests}
\label{sect:numerics}
We present now several numerical tests, which illustrate the theoretical results obtained in Section~\ref{sect:FP}. In particular, they show that the large time numerical solution to the Boltzmann-type equation with cutoff~\eqref{eq:Boltzmann} is consistently approximated, for $\epsilon>0$ small, by either stationary distribution~\eqref{eq:finf.n=1},~\eqref{eq:finf.n=2} depending on the assumed model of binary interactions.

\begin{algorithm}[!t]
	\caption{Nanbu-Babovsky Monte Carlo scheme with rejection for~\eqref{eq:Boltzmann}}
	\begin{algorithmic}[1]
		\STATE fix $N>1$ (number of particles, even) and $\Delta{t}\in (0,\,\epsilon]$ (time step)
		\STATE sample $N$ particles from the initial distribution $f^0$; let $\{s_i^0\}_{i=1}^{N}$ be their microscopic states
		\FOR{$\ell=0,\,1,\,2,\,\dots$}
			\STATE set $\tilde{N}:=\frac{\Delta{t}}{\epsilon}N$
			\STATE \label{alg:sample} sample uniformly $\frac{\tilde{N}}{2}$ pairs of indexes $(i,\,j)$ with $i,\,j\in\{1,\,\dots,\,N\}$, $i\neq j$ and no repetition
			\FOR{every sampled pair $(i,\,j)$}
				\STATE let $s_i':=s_i^\ell+I(s_i^\ell,\,s_j^\ell)+{(s_i^\ell)}^\delta\eta$, cf.~\eqref{eq:interaction_micro}, with $\epsilon$-scaled $I$, $\eta$ (quasi-invariant regime)
				\IF{$s_i'\geq 0$} \label{alg:rej.1}
					\STATE set $s_i^{\ell+1}:=s_i'$
				\ELSE
					\STATE set $s_i^{\ell+1}:=s_i^\ell$
				\ENDIF \label{alg:rej.2}
				\STATE set $s_j^{\ell+1}:=s_j^\ell$
			\ENDFOR
			\STATE set $s_i^{\ell+1}:=s_i^\ell$ for all indexes $i$ which were not sampled in step~\ref{alg:sample}
		\ENDFOR
	\end{algorithmic}
	\label{alg:nanbu}
\end{algorithm}

For the numerical solution of the Boltzmann-type equation with cutoff~\eqref{eq:Boltzmann}, we adopt a direct simulation Monte Carlo (MC) method. We refer the interested reader to~\cite{pareschi2001ESAIMP,pareschi2013BOOK} for an introduction. Here, we simply report an essential algorithm which implements an MC scheme suited to our equation, see Algorithm~\ref{alg:nanbu}. In particular, unlike standard MC algorithms, we take into account that some binary interactions may need to be rejected, if they produce negative post-interaction headways (see lines~\ref{alg:rej.1} to~\ref{alg:rej.2} in Algorithm~\ref{alg:nanbu}). It is worth remarking that, besides updating the microscopic states of the particles with the MC scheme, we also need to reconstruct their probability density function at every time step. For this, we recall that several approaches are possible, such as e.g., standard histograms (which we use in this paper), the weighted area rule or kernel density estimation-type strategies.

In the following tests, we invariably use a sample of $N=10^5$ particles. Moreover, for density reconstruction purposes, we take $s$ in a bounded interval $[0,\,S]\subset\R_+$ and we discretise the latter by means of a certain number $N_S$ of grid points. In particular, for the model with $n=1$ we use $S=20$ and $N_S=200$, while for the model with $n=2$ we use $S=10$ and $N_S=100$.

\subsection{Log-normal equilibrium (\texorpdfstring{$\boldsymbol{n=1}$}{})}
\label{sect:num_lognorm}
\begin{figure}[!t]
\centering
\includegraphics[scale=0.43]{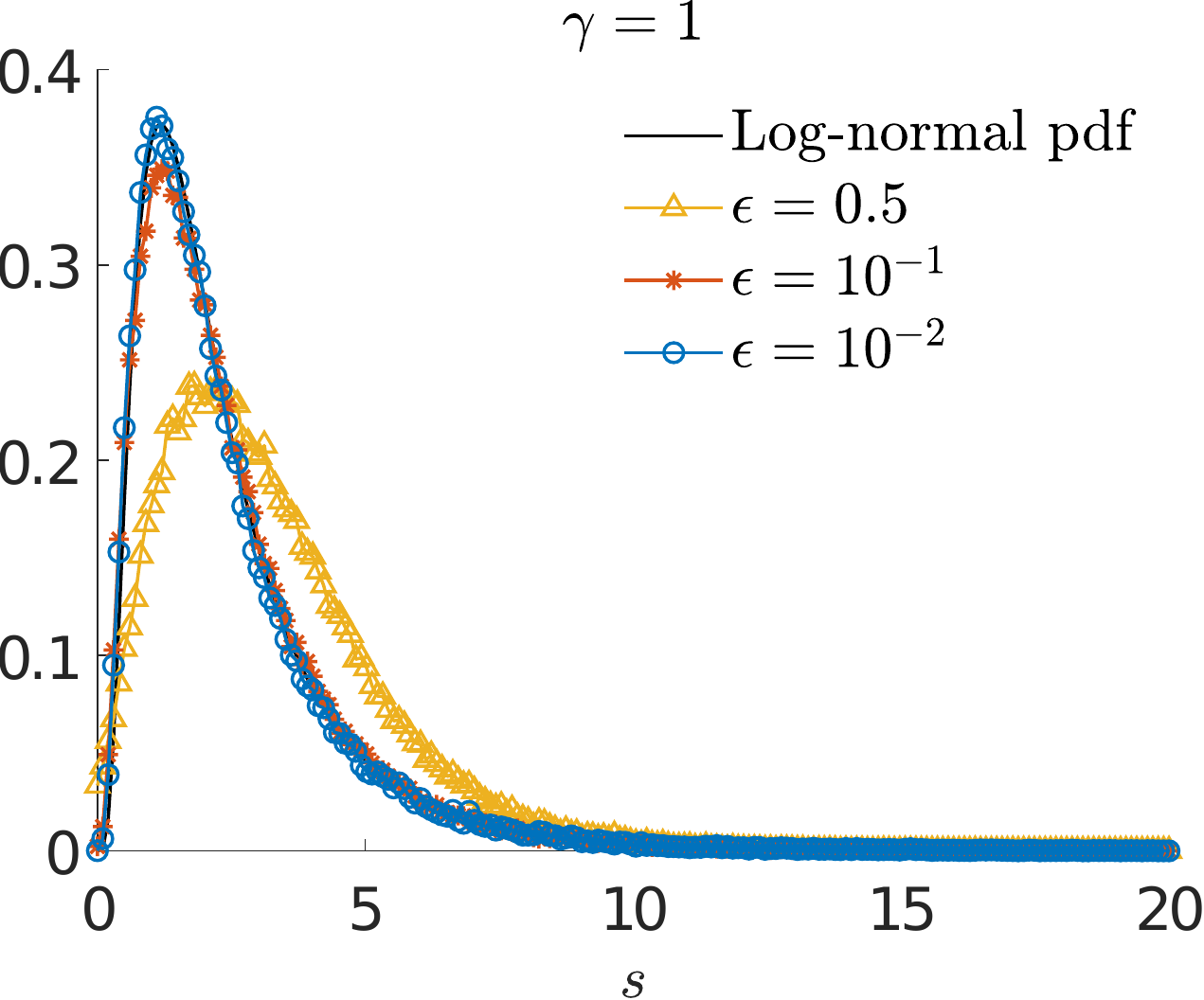} \qquad
\includegraphics[scale=0.43]{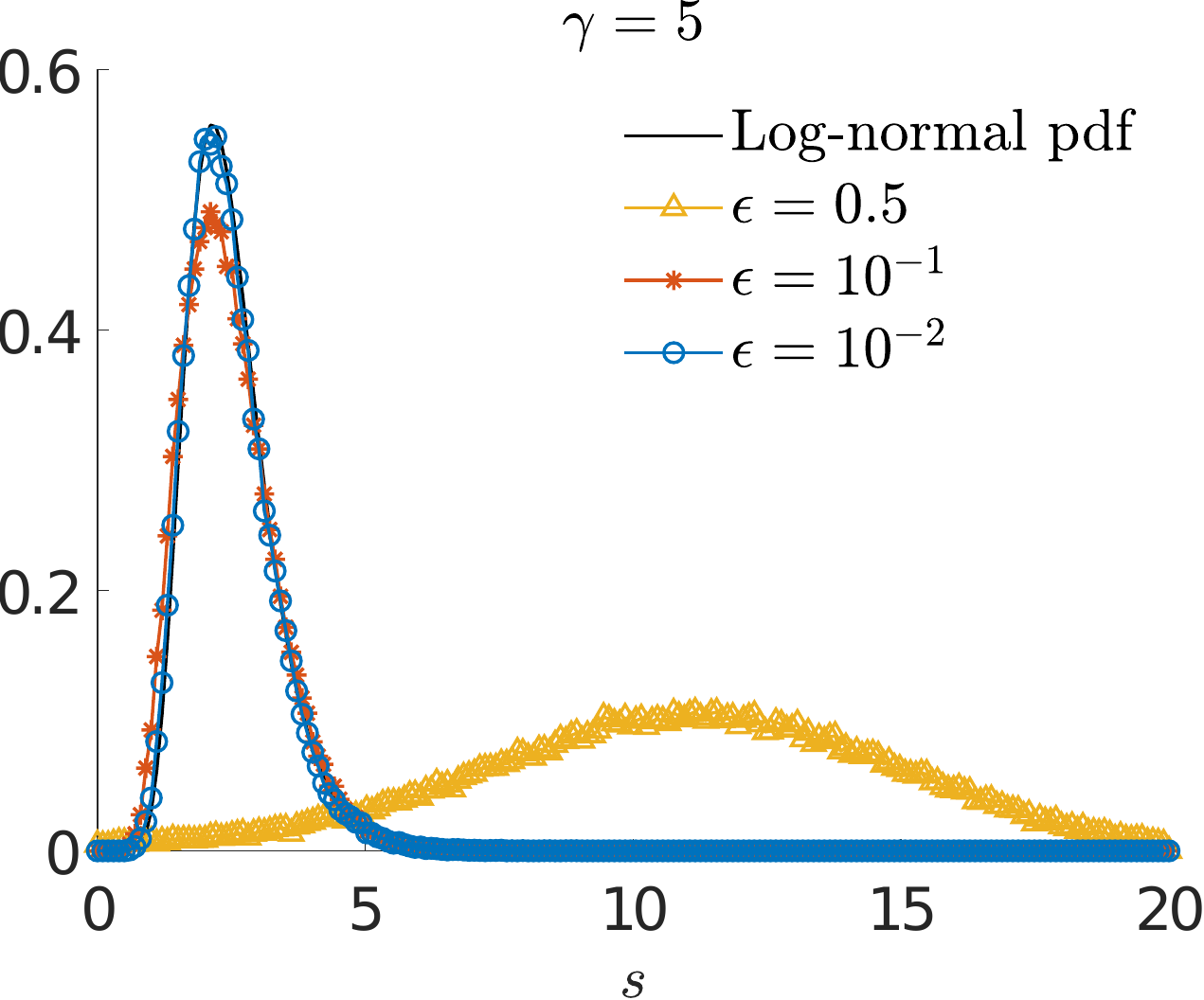}
\caption{Follow-the-Leader model with $n=1$. Comparison of the large time numerical solution of~\eqref{eq:Boltzmann} with the Fokker-Planck equilibrium distribution~\eqref{eq:finf.n=1} for a decreasing scaling parameter $\epsilon$ and two different values of the parameter $\gamma$ in~\eqref{eq:binary.n=1}.}
\label{fig:n=1.1}
\end{figure}

We consider first the binary interaction scheme~\eqref{eq:binary.n=1} with $\delta=\frac{1}{2}$ and the quasi-invariant scaling~\eqref{eq:quasi-inv.n=1}. In particular, we take for $\eta$ a centred uniform law, so as to meet Assumption~\ref{ass:logs.Y}\ref{ass.item:Y.symm}. Moreover, we prescribe the following initial condition:
\begin{equation}
	f(s,\,0)=
		\begin{cases}
			\frac{1}{5} & \text{if } 0\leq s\leq 5 \\
			0 & \text{otherwise},
		\end{cases}
	\label{eq:init_state}
\end{equation}
whence the mean headway is initially $h=\frac{5}{2}$. In Figure~\ref{fig:n=1.1}, we show the numerical solution of~\eqref{eq:Boltzmann} in the scaled regimes $\epsilon=0.5,\,10^{-1},\,10^{-2}$ obtained with Algorithm~\ref{alg:nanbu} after $T=20$ time steps. A direct comparison with the log-normal equilibrium distribution~\eqref{eq:finf.n=1}, also plotted in Figure~\ref{fig:n=1.1}, confirms that if $\epsilon$ is sufficiently small ($\epsilon=O(10^{-2})$ in this case) the Fokker-Planck asymptotics provides a consistent approximation of the large time Boltzmann-type solution. Conversely, if $\epsilon$ is not small enough, the large time Boltzmann-type solution may differ consistently from the Fokker-Planck equilibrium (cf. e.g., the case $\epsilon=0.5$). One of the main reasons is that when $\epsilon$ is large many interactions produce $s'<0$ and are therefore discarded by the collision kernel $\chi(s'\geq 0)$. Consequently, the statistical description provided by~\eqref{eq:Boltzmann} is considerably different from that provided by~\eqref{eq:FP-strong.n=1}.

\begin{figure}[!t]
\centering
\includegraphics[scale=0.43]{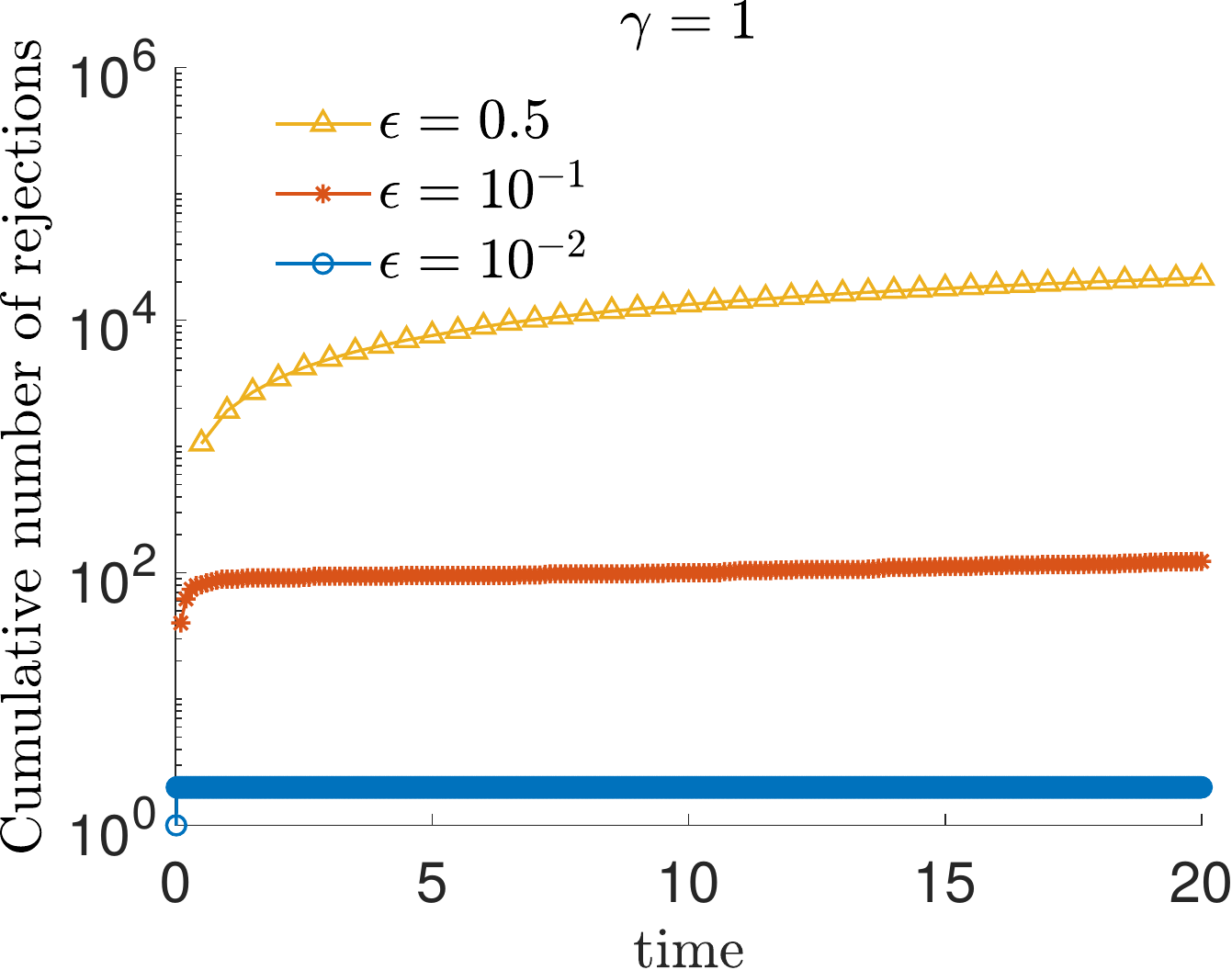} \qquad
\includegraphics[scale=0.43]{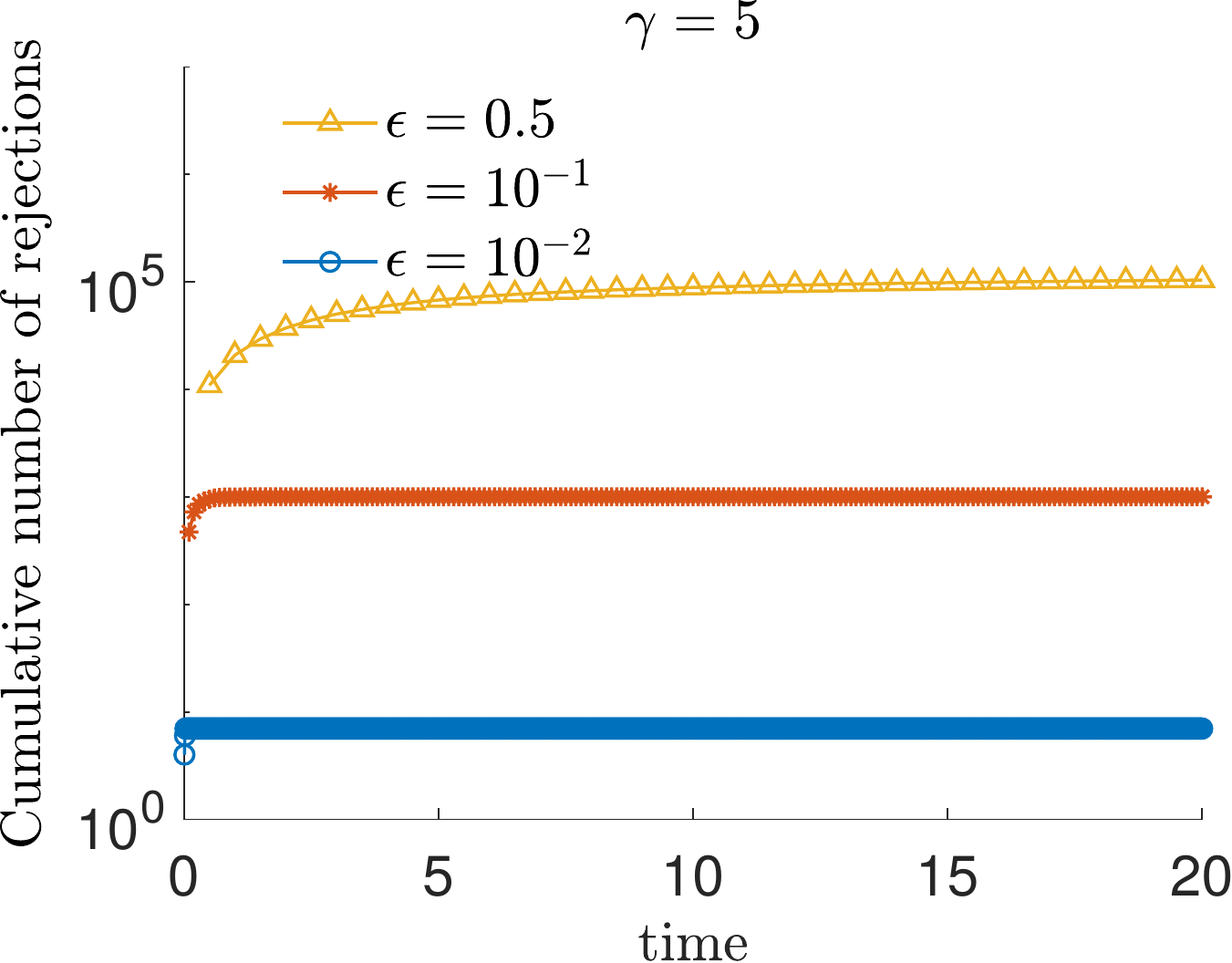}
\caption{Follow-the-Leader model with $n=1$. Cumulative number of particles rejected by the MC algorithm~\ref{alg:nanbu} in time (semi-logarithmic scale).}
\label{fig:n=1.2}
\end{figure}

To further investigate the latter aspect, we track the cumulative number of rejections performed by the MC algorithm~\ref{alg:nanbu}. In Figure~\ref{fig:n=1.2}, we show the evolution of such a number in time, starting from the initial condition~\eqref{eq:init_state}. We observe that, when $\epsilon$ is small enough, this number remains constant in time, which indicates that the binary interactions tend to produce only physically acceptable microscopic states. The non-zero cumulative number of rejections is simply due to the arbitrarily chosen initial condition, as the jump at $t=0$ in the curve for $\epsilon=10^{-2}$ clearly shows.


\subsection{Gamma equilibrium (\texorpdfstring{$\boldsymbol{n=2}$}{})}
We repeat the same tests as in Section~\ref{sect:num_lognorm} for the binary interaction scheme~\eqref{eq:binary.n=2} with $\delta=\frac{1}{2}$ under the quasi-invariant scaling~\eqref{eq:quasi-inv.n=2}. Hence, we compare the large time numerical solution of the Boltzmann-type equation~\eqref{eq:Boltzmann} with the gamma equilibrium distribution~\eqref{eq:finf.n=2} of the Fokker-Planck equation~\eqref{eq:FP-strong.n=2} obtained in the quasi-invariant limit.

\begin{figure}[!t]
\centering
\includegraphics[scale=0.43]{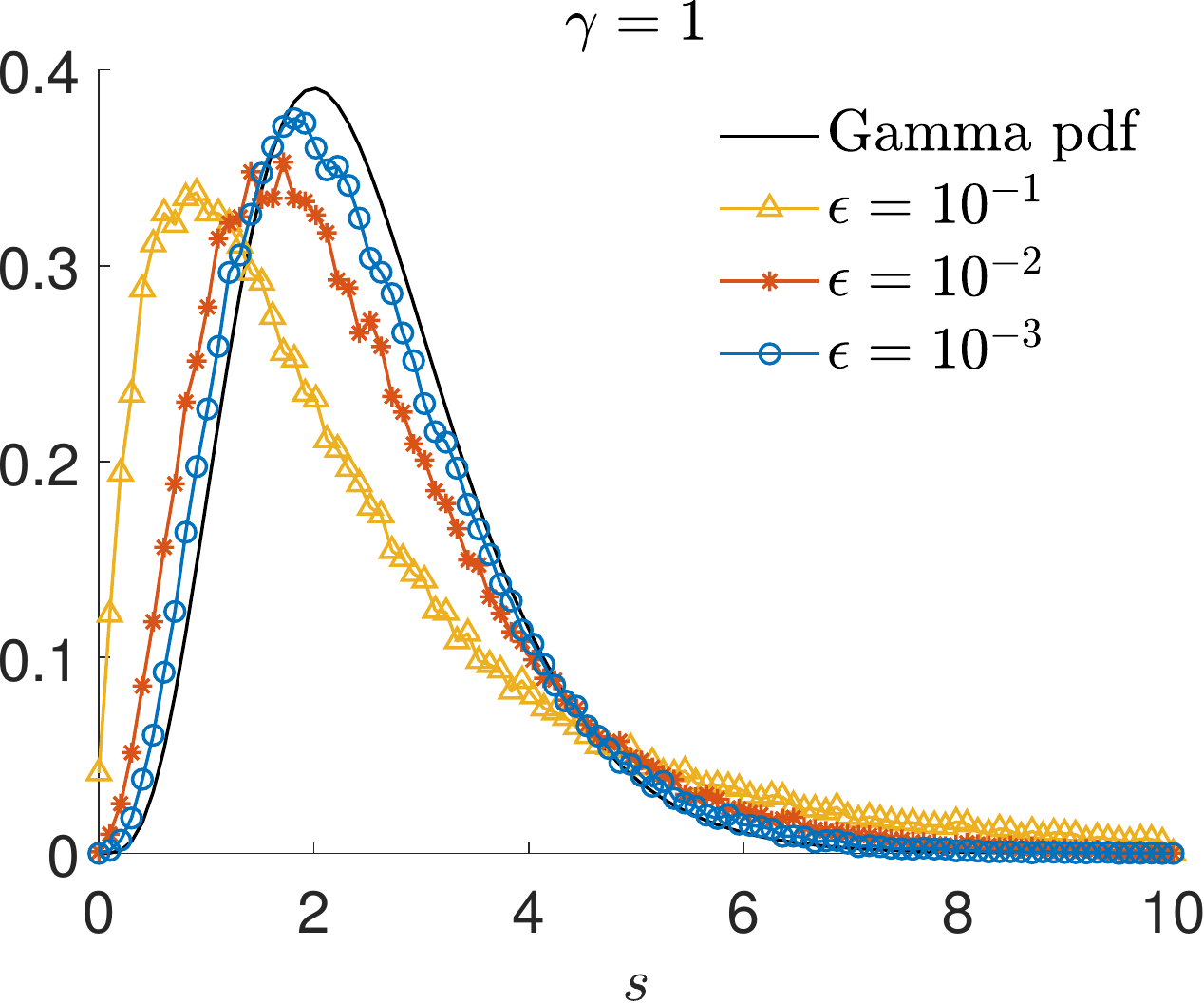} \qquad
\includegraphics[scale=0.43]{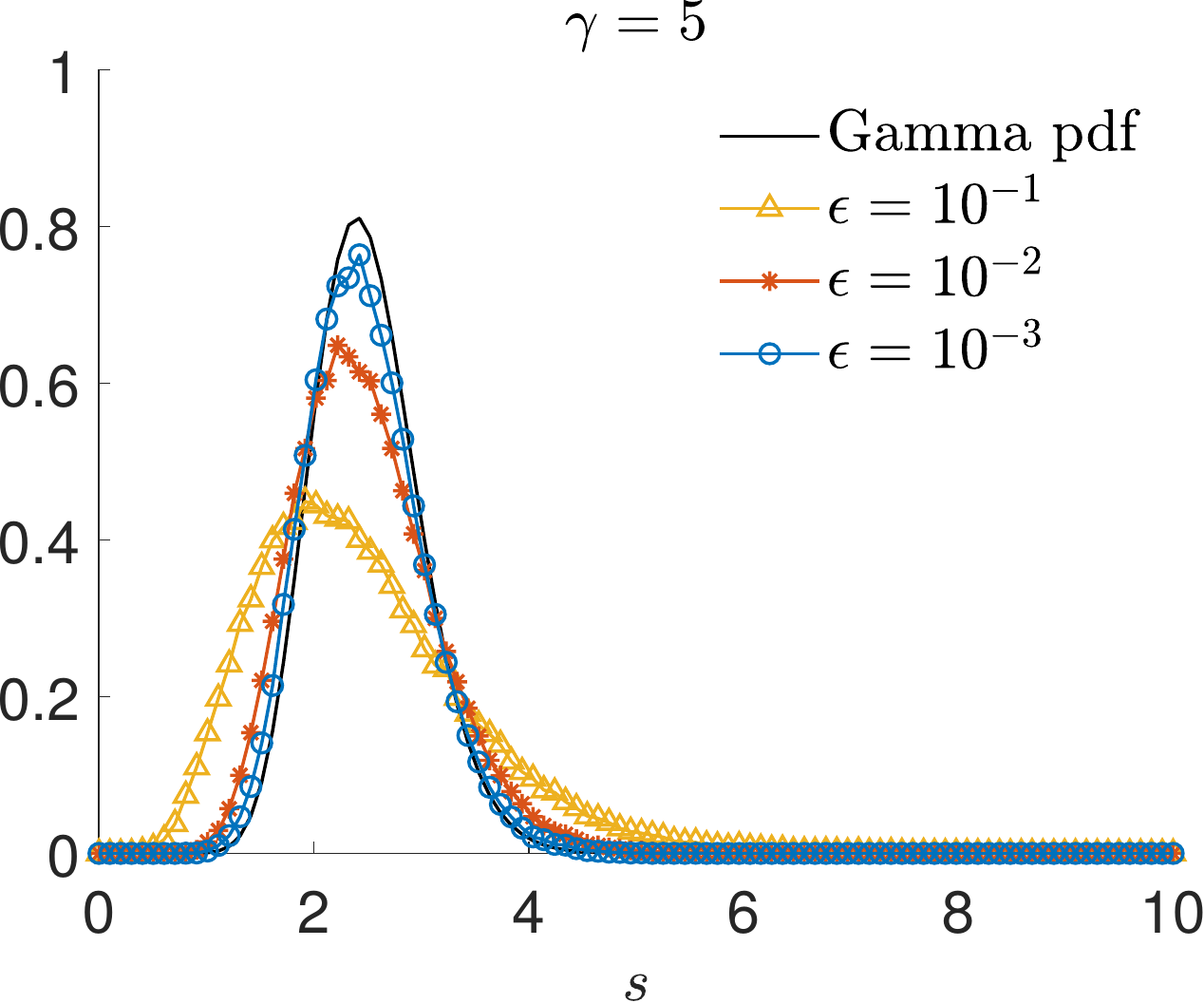}
\caption{Follow-the-Leader model with $n=2$. Comparison of the large time numerical solution of~\eqref{eq:Boltzmann} with the Fokker-Planck equilibrium distribution~\eqref{eq:finf.n=2} for a decreasing scaling parameter $\epsilon$ and two different values of the parameter $\gamma$ in~\eqref{eq:binary.n=2}.}
\label{fig:n=2.1}
\end{figure}

\begin{figure}[!t]
\centering
\includegraphics[scale=0.43]{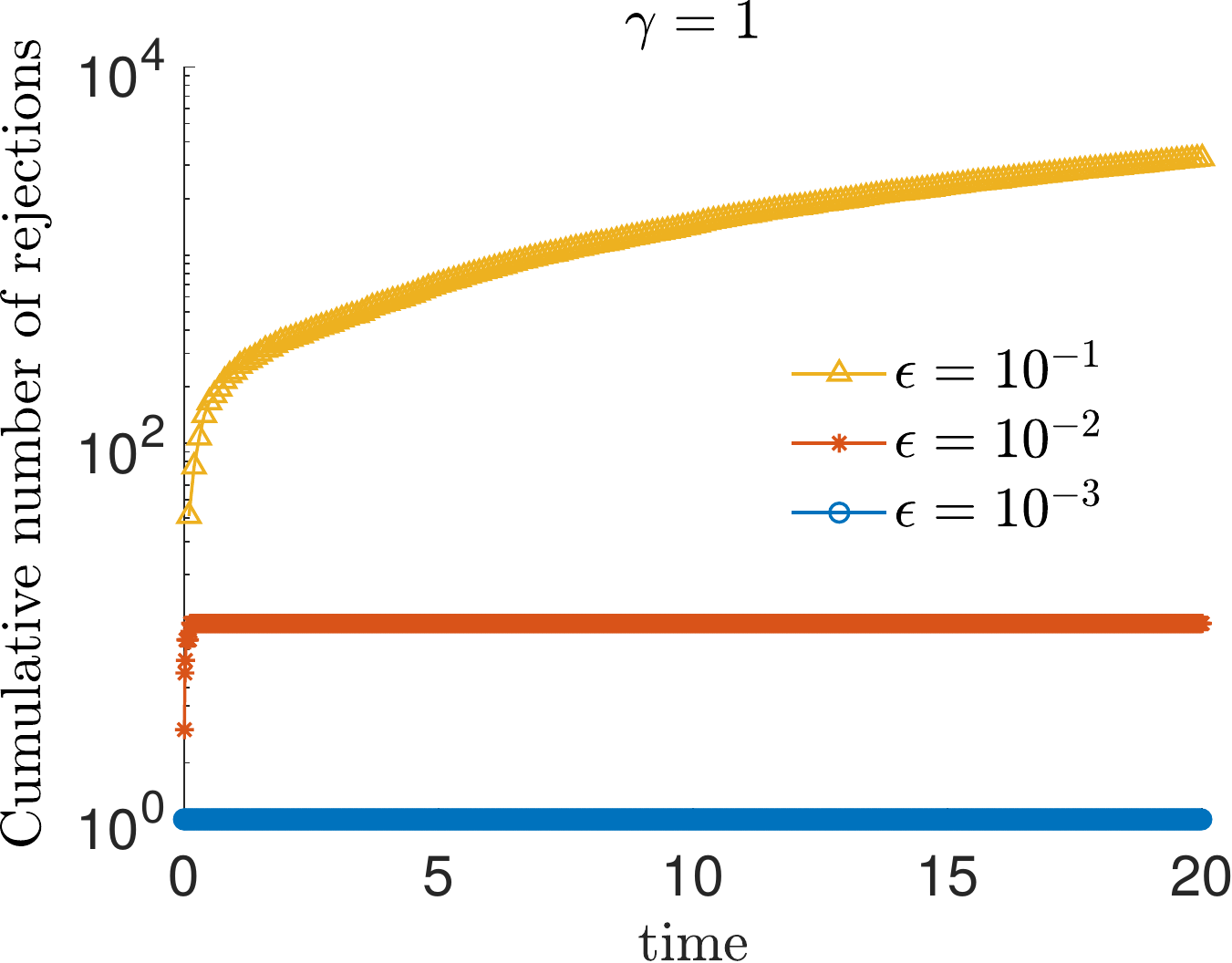} \qquad
\includegraphics[scale=0.43]{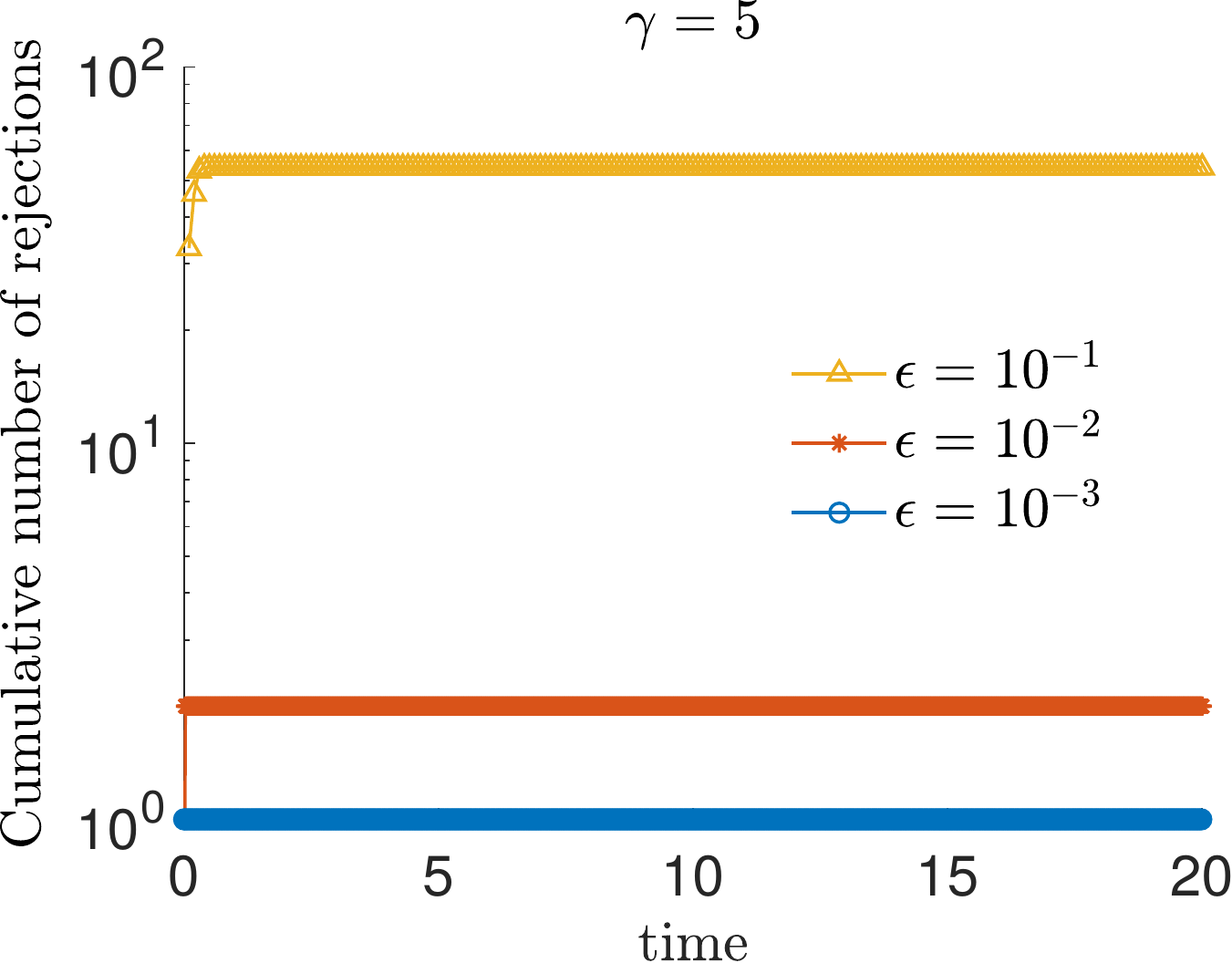}
\caption{Follow-the-Leader model with $n=2$. Cumulative number of particles rejected by the MC algorithm~\ref{alg:nanbu} in time (semi-logarithmic scale).}
\label{fig:n=2.2}
\end{figure}

Figure~\ref{fig:n=2.1} confirms that, for $\epsilon$ sufficiently small ($\epsilon=O(10^{-3})$ in this case), the large time Boltzmann solution approaches consistently the analytical Fokker-Planck equilibrium. Moreover, Figure~\ref{fig:n=2.2} shows that, for decreasing $\epsilon$, the cumulative number of rejections performed by the MC algorithm~\ref{alg:nanbu} diminishes and remains constant in time.


\section{Conclusions}
\label{sect:conclusions}
In this paper we have shown that a Boltzmann-type kinetic approach may be successfully applied to Follow-the-Leader (FTL) traffic models to explain the emergence of various statistical distributions used to interpolate empirical traffic data. Specifically, we have recovered the log-normal and the gamma profiles of the headway and time headway distributions from FTL models of the form
$$	\begin{cases}
		\dot{x}_i=v_i \\
		\dot{v}_i=a{\left(\dfrac{v_i}{x_{i+1}-x_i}\right)}^n(v_{i+1}-v_i)
	\end{cases} $$
with $a>0$ and $n=1,\,2$, respectively.

The further inclusion of stochastic fluctuations at the level of microscopic vehicle interactions, modelling the random behaviour of the drivers superimposed to the purely deterministic FTL dynamics, has turned out to be a crucial point. Indeed, the type of stationary distribution resulting from the kinetic model depends on the rate at which energy is introduced in the system by the interactions. We have described the stochastic fluctuations by means of a term of the form $s^\delta\eta$, where $s\geq 0$ is the headway, $\delta>0$ is a parameter and $\eta\in\R$ is a centred random variable with non-zero variance. In this setting, the input rate of the energy is $s^\delta$, which increases with $s$ to model the fact that for close vehicles the deterministic FTL dynamics dominate over the stochastic fluctuations while for far apart vehicles the converse holds. The log-normal and gamma distributions have been obtained for $\delta=\frac{1}{2}$. Conversely, still in the case $n=2$, we have shown that for $\delta=1$ an inverse gamma distribution is obtained, which belongs to the class of fat tailed distributions sometimes also cited in the experimental literature.

From the technical point of view, treating the cases with $\delta=\frac{1}{2}$ has required to deal with ``collisional'' models with cutoff. This means that in the Boltzmann-type equation we have considered a non-constant collision kernel of the form $\chi(s'\geq 0)$, where $\chi$ denotes the characteristic function and $s'$ is the post-interaction headway. Such a kernel discards from the statistical description of the system possible interactions leading to unphysical negative headways and turns out to be necessary because for $\delta=\frac{1}{2}$ it is impossible to rule out \textit{a priori} such interactions. On the other hand, for $\delta=1$ a more standard Maxwellian description may be adopted, because \textit{a priori} bounds on $\eta$ and the parameters of the interactions can be established which guarantee the non-negativity of the post-interaction headway.

The analytical determination of the stationary distributions mentioned above has been possible in a particular regime of the microscopic parameters, called the quasi-invariant regime. Essentially, it corresponds to the case in which each vehicle interaction produces a very small variation of the headway but the interaction frequency is very high. In this sense, it is reminiscent of the grazing collision regime of the classical kinetic theory. In such a regime, the Boltzmann-type equation can be consistently approximated by a Fokker-Planck equation, which is more amenable to analytical investigations including the possible explicit computation of the large time distributions. Nevertheless, the application of this theory to kinetic models with cutoff is non-standard and has represented the main difficulty to overcome in this paper from both the analytical and the numerical points of view.

We believe that the techniques discussed in this paper may further foster the application of kinetic theory methods to new problems in the wide realm of multi-agent systems, which for various reasons may require non-constant interaction kernels, see e.g.,~\cite{furioli2020M3AS,tosin2019MCRF_preprint}, and whose investigation might have been partly discouraged so far by the lack of proper analytical and numerical tools.

\section*{Acknowledgements}
This research was partially supported  by the Italian Ministry for Education, University and Research (MIUR) through the ``Dipartimenti di Eccellenza'' Programme (2018-2022), Department of Mathematical Sciences ``G. L. Lagrange'', Politecnico di Torino (CUP: E11G18000350001) and Department of Mathematics ``F. Casorati'', University of Pavia; and through the PRIN 2017 project (No. 2017KKJP4X) ``Innovative numerical methods for evolutionary partial differential equations and applications''.

This work is also part of the activities of the Starting Grant ``Attracting Excellent Professors'' funded by ``Compagnia di San Paolo'' (Torino) and promoted by Politecnico di Torino.

Both authors are members of GNFM (Gruppo Nazionale per la Fisica Matematica) of INdAM (Istituto Nazionale di Alta Matematica), Italy.

\bibliographystyle{plain}
\bibliography{TaZm-Boltzmann_cutoff_traffic}

\end{document}